\documentclass[english,twocolumn,aps,superscriptaddress,prx,eqsecnum]{revtex4-1}
\usepackage{babel}
\usepackage{amsmath}
\usepackage{amssymb}
\usepackage{graphicx,color}
\usepackage{epsfig,natbib}
\newcommand{\be}{\begin{equation}}
\newcommand{\ee}{\end{equation}}
\newcommand{\bea}{\begin{eqnarray}}
\newcommand{\eea}{\end{eqnarray}}

\begin{document}
\title{Phase diagrams of Majorana-Hubbard ladders}

\author{Armin Rahmani}
\affiliation{Department of Physics and Astronomy and Advanced Materials Science and Engineering Center, Western Washington University, Bellingham, Washington 98225, USA}
\author{Dmitry Pikulin}
\affiliation{Microsoft Quantum, Microsoft Station Q, University of California, Santa Barbara, California 93106-6105, USA}
\author{Ian Affleck}
\affiliation{Department of Physics and Astronomy and Stewart Blusson Quantum Matter Institute, University of British Columbia, 
Vancouver, B.C., Canada, V6T1Z1}

\begin{abstract}
Models of interacting Majorana modes may be realized in vortex lattices in superconducting films in contact with topological insulators and may be tuned to 
the strong-interaction regime by adjusting the chemical potential. Extending the results on  one- and two-dimensional Majorana-Hubbard models, here we determine the phase diagrams of two- and four-leg ladders 
using both field theory and the {density-matrix renormalization group (DMRG)} methods.
\end{abstract}
\maketitle
\section{Introduction}

Majorana zero modes (MZM) are expected to emerge in various superconducting systems~[\onlinecite{Kitaev2001, Fu2008, Lutchyn2010, Oreg2010}]. Motivated by the promise of topological computing, there has been significant experimental progress in their realizations recently~[\onlinecite{Konig2007, Zhang2009, Xia2009, Mourik2012, lutchyn2018majorana}].
In some of the proposed realizations, multiple MZMs with tunable couplings between them can arise~[\onlinecite{Fu2008, Zhou2013, Biswas2013, Chiu2015, Liu2015, Pikulin2015}]. From a fundamental viewpoint, Majorana fermions are the real (with Hermitian creation/annihilation operators) counterparts of complex (with non-Hermitian creation/annihilation operators) fermions such as ordinary electrons. Electrons are elementary particles in all materials and their interplay and interactions gives rise to several interesting phases of matter. With the rapid progress in realizing emergent Majorana fermions, understanding the many-body phases of matter that emerge from the interplay of many Majorana fermions as their elementary building blocks is a natural and timely extension in condensed matter physics~[\onlinecite{Hermanns2014,Chiu2015, Pikulin2015, Rahmani2015, Rahmani2015a, Milsted2015, Witczak2016, Ware2016,Gangadharaiah2011,Lobos2012, li2018majorana}].

Some of the most interesting phenomena in many-body electron systems occur due to strong electron-electron interactions. These include high-temperature superconductors and fractional quantum Hall liquids, which unlike simple metals and insulators, cannot be understood in terms of simple weakly interacting electrons. Interacting many-body Majorana systems are of particular interest because at least in one of their physical realizations, the strongly interacting regime can be accessed in a tunable way, even in the absence of strong interactions between the underlying electrons.

The above-mentioned particular realization is a superconducting film in contact with a topological insulator in the presence of a transverse magnetic field is predicted to have a Majorana mode 
localized in every vortex core~[\onlinecite{Fu2008}]. The low-energy effective Hamiltonian contains short-range hopping and interaction terms.  By tuning the chemical 
potential to zero with respect to the Dirac point of the topological insulator surface, the hopping term can be made to vanish due to an extra chiral symmetry, which changes the topological classification from $Z_2$ to $Z$~[\onlinecite{Teo2010}]. Consequently, tuning the chemical potential to small values results in small hopping parameters and brings the system into the strong-coupling regime~[\onlinecite{Chiu2015}].

The simplest one-dimensional 
case, i.e., the Majorana-Hubbard chain, was studied using a combination of field theory and density-matrix renormalization group (DMRG) methods~[\onlinecite{Rahmani2015,Rahmani2015a}]. In analogy with the Hubbard model, the shortest range, most local possible interactions are kept in the model. Since for a Majorana mode, $\gamma^2=1$, the most local interaction for a one-dimensional chain involves four consecutive sites on the chain.  The Majorana-Hubbard chain was shown to have a rich phase diagram, which includes a supersymmetric tricritical Ising phase transition and strong-coupling phases 
where the Majorana modes combine in pairs to form ordinary complex (``Dirac'') fermions, spontaneously breaking various discrete symmetries of the model. Other slightly less local one-dimensional models, which allow for extended interactions between four Majorana modes in clusters of five consecutive sites exhibit a similar phase diagram with the phase transition at weaker interaction strengths~[\onlinecite{Hosho,Obrien}].

The two-dimensional (2D) square lattice version of this model, with Hamiltonian,
\be 
\begin{split}
H=\sum_{m,n}\big\{& it\gamma_{m,n}[(-1)^n\gamma_{m+1,n}+\gamma_{m,n+1}]
\\&+g\gamma_{m,n}\gamma_{m+1,n}\gamma_{m+1,n+1}\gamma_{m,n+1}\big\}\end{split}\label{eq:H2d}
\ee
was studied in Ref. [\onlinecite{Affleck2017}] where similar strong-coupling phases were argued to occur using mean-field theory, field theory and renormalization group (RG)
methods. Due to the presence of square plaquettes, the most local interaction involves four sites around a square plaquette in the Majorana-Hubbard model on the square lattice. The infinite coupling limit was studied in Ref. [\onlinecite{Kamiya}] using quantum Monte Carlo and an exact mapping into the ``Compass'' spin model, 
obtaining results largely consistent with Ref. [\onlinecite{Affleck2017}]. The Majorana-Hubbard model has also been studied on the honeycomb lattice~\cite{li2018majorana} (with the most local interaction containing lattice sites and their three nearest neighbors),  using mean-field theory and exact diagonalization.

In the field of strongly correlated electrons, ladders play an important role. Various models of interacting fermions and spins have been studied on ladders. They are of considerable interest in their own right as quasi-one-dimensional models as the natural extensions of one-dimensional systems, which can exhibit novel phases and phase transitions. Furthermore, developments in DMRG have turned ladders into a useful tool for studying the physics of two-dimensional problems, which are not amenable to numerical quantum Monte Carlo simulations due to the sign problem. By systematically pushing the DMRG calculations to wider and wider ladders, one may reach a limit, where the essential physics of the two-dimensional models can be inferred by extrapolation. This is particularly useful because, unlike some approximations we need to resort to in studying two-dimensional model, e.g., various mean-filed theories, DMRG is numerically accurate and controlled.

In the present paper, as the first study of Majorana-Hubbard ladders, we focus on two- and four-leg ladders with square plaquettes. We find rich phase diagram, which have similar features to both the chain and the two-dimensional square lattice, and several unique features. We show that the two-leg
case maps into a well-understood model, the XXZ $S=1/2$ chain. The four-leg case exhibits novel behavior, which we study with 
a combination of field theory and {DMRG }methods. We find 
that, in analogy with the Hubbard-Majorana chain and 2D square lattice, Majorana modes combine to form occupied/empty Dirac modes at strong coupling, breaking discrete symmetries, in both two- and four-leg cases. Several phase transitions occur in the four-leg ladder between various critical phases with central charges $c=1, 3/2, 2$. A theoretical description is provided for two of these phase transitions, which is supported by numerical results. The first transition upon increasing the interaction strength from the noninteracting point is found to be of the Kosterlitz-Thouless (Lifshitz) type  for $g>0$ ($g<0$) .

The outline of the remainder of the paper is as follows. In Sec. II, we review results on the 2D case. In Sec. III, we study the two-leg ladder, with periodic boundary conditions (PBC), showing that it has a hidden U(1) symmetry 
and can be mapped into the XXZ $S=1/2$ spin chain model.  It has a massless phase at weak coupling with broken symmetry states occurring at sufficiently strong coupling of either sign, 
 corresponding to Majoranas on rungs forming Dirac fermions, consistent with our predictions in the 2D case.
  In Sec. IV, we show that the weak coupling limit for a ladder with an even number of legs and PBC 
is equivalent to the two-leg case, with the corresponding massless phase. We then study the strong-coupling limit of the four-leg ladder with PBC, showing that Majorana pairing occurs, 
largely consistent with our predictions \cite{Affleck2017} in D=2.   In Sec. V, we study the two and four-leg ladder with 
open boundary conditions in the $y$ direction, which lead to quite different behavior. In Sec. VI, we include second
 neighbor interactions for the two-leg case. 
In Sec. VII, we map the four-leg Majorana ladder into a two-leg spin ladder and check that the strong-coupling limit discussed 
in Sec. IV is recovered. In Sec. VIII, we analyze analytically the phase transition that occurs in the four-leg ladder for $g>0$.  In Sec. IX, we present numerical results on the phase diagram of the four-leg ladder. 
Section X contains conclusions.

\begin{figure}
\epsfxsize=7 cm
\includegraphics*[width=1.0\columnwidth]{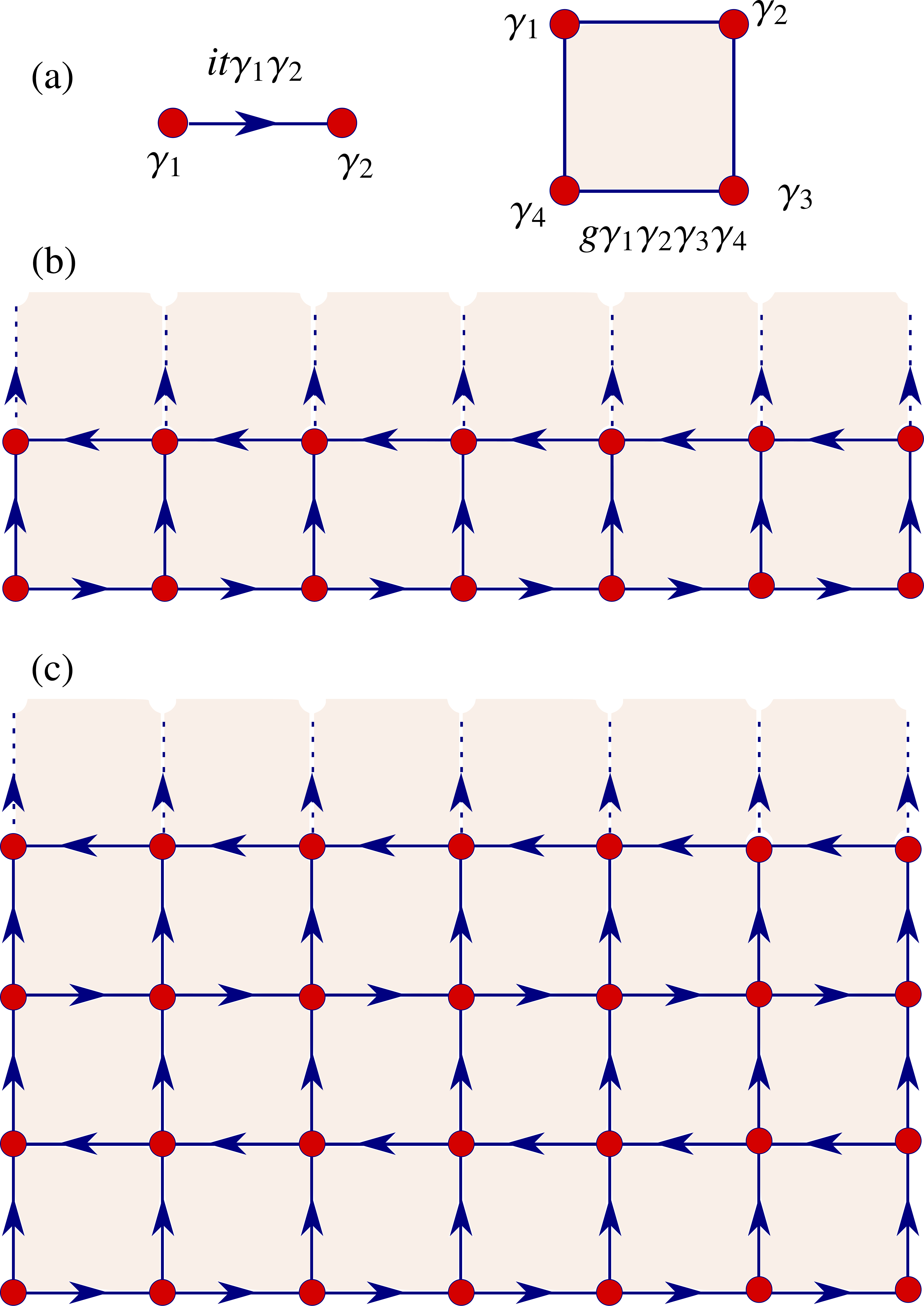}
\caption{(a) The convention for the hopping and interactions terms along bonds and plaquettes. (b) The two-leg ladder with periodic boundary conditions in the $y$ direction. The periodic boundary conditions double the interaction strength and remove the vertical hopping compared with open boundary conditions. (c) The four-leg ladder with periodic boundary conditions.}
\label{fig:schem}
\end{figure}

\section{Review of 2D case}
\begin{figure}
\epsfxsize=8 cm
\includegraphics*[width=1.0\columnwidth]{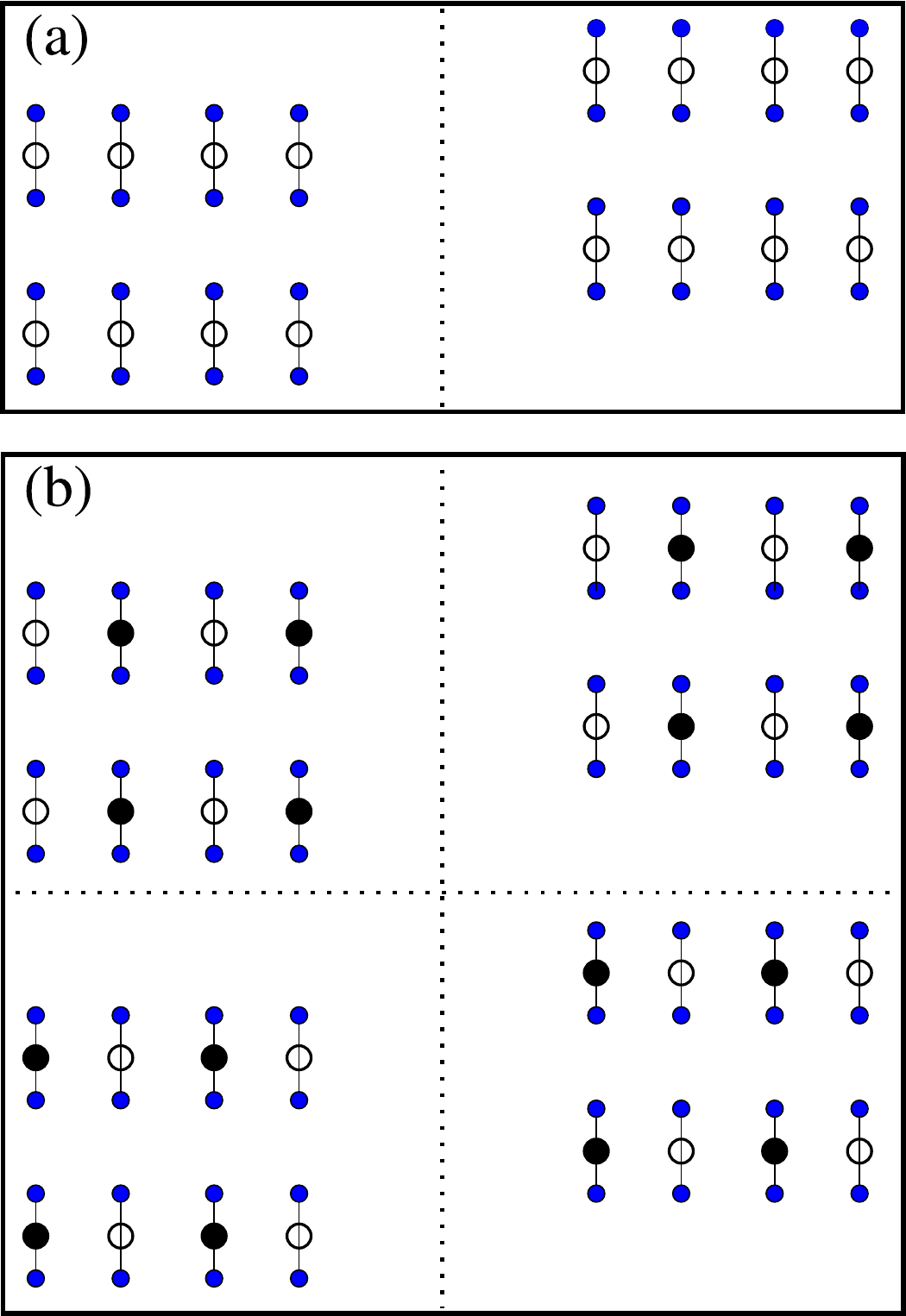}
\caption{(a) The symmetry-breaking pattern for two of the four strong-coupling ground states of the two-dimensional Majorana-Hubbard model for $g>0$ on the square lattice predicted in Ref. [\onlinecite{Affleck2017}] using mean-field theory. The other two states can be obtained by a $\pi/2$ rotation. (b) Four of the eight symmetry-breaking patterns of the mean-field strong-coupling ground states predicted in Ref. [\onlinecite{Affleck2017}] for $g<0$. The other four states can be obtained by a $\pi/2$ rotation. Blue circles are the Majorana modes, a bond between them indicates the combination of the MZMs into a Dirac fermion. The larger circle on the bond represent the occupation of the Dirac mode.}
\label{fig:2d}
\end{figure}

Due to the alternating sign of the horizontal hopping term in Eq.~\eqref{eq:H2d} (stemming from the flux quantization condition in the underlying vortex lattice~\cite{Stern2008}), while translation by one lattice spacing in the $x$ direction is a symmetry, in the $y$ direction, only translation by two lattice spacings is a symmetry. Thus,  the unit cell contains two sites on the same column and it convenient to introduce $e/o$ labels for even and odd rows:
\be \gamma_{2j}\equiv \gamma^e_{2j},\ \  \gamma_{2j+1}\equiv \gamma^o_{2j+1}.\ee
(Here we use slightly different notation than in Ref. [\onlinecite{Affleck2017}] to simplify some formulas.) For a chain of $2W$ rows of length $L$ we Fourier transform the Majorana operators as
\bea \gamma^e_{\vec k}&\equiv & {1\over \sqrt{2WL}}\sum_{m,n}e^{i(mk_x+2nk_y)}\gamma^e_{m,2n}, \\
\gamma^o_{\vec k}&\equiv &  {1\over \sqrt{2WL}}\sum_{m,n}e^{i[mk_x+(2n+1)k_y]}\gamma^0_{m,2n+1}.
\eea
The hopping term in $H$ then becomes:
\be \begin{split}
H_0=-4t\sum_{k_x>0,k_y}\Big[&\left(\gamma^{e\dagger}_{\vec k}\gamma^e_{\vec k}-\gamma^{o\dagger}_{\vec k}\gamma^o_{\vec k}\right)\sin k_x\\
&+\left(\gamma^{e\dagger}_{\vec k}\gamma^o_{\vec k}+\gamma^{o\dagger}_{\vec k}\gamma^e_{\vec k}\right)\sin k_y\Big] .
\end{split}\label{H0}\ee
Here we have used the fact that $\gamma^{e/o}_{-\vec k}=\gamma^{e/o\dagger }_{\vec k}$ to restrict the Brillouin zone to $0\leq k_x<\pi$, $-\pi /2\leq k_y<\pi /2$. 
Diagonalizing the above noninteracting Hamiltonian gives the following energy bands:
\be E_{\pm}=\pm 4t\sqrt{\sin^2 k_x+\sin^2k_y}.\label{E0}\ee
The low energy Hamiltonian corresponds to two two-component relativistic Majorana fermions at the two ``Dirac points'' $(0,0)$ and $(\pi ,0)$, which can be combined into 
a single relativistic Dirac fermion, $\psi$.  The interaction term becomes:
\be {\cal H}_{int}=32g(\bar \psi \psi )^2,\ee
which is an irrelevant interaction in the relativistic $(2+1)$-dimensional field theory, leading to a massless phase for sufficiently weak coupling. We predicted in Ref.~[\onlinecite{Affleck2017}] that,  at a critical 
positive coupling, $g_{c4}$, there is a transition into a phase with pairs of neighboring Majoranas forming Dirac fermions. At a mean-field level, these 
Dirac energy levels are either filled or empty as indicated in Fig. (\ref{fig:2d}); unfilled circles correspond to empty states. 
 In addition to these ground states two others occur, rotated by $\pi /2$ with 
Dirac fermions forming on horizontal links.  For large enough negative $g$, $g<g_{c1}$ a symmetry-breaking phase occurs with the Dirac fermions levels alternating filled and empty 
as indicated in Fig. (\ref{fig:2d}). As shown in Fig.  (\ref{fig:2d}), the strongly coupled ordered phase is four-fold (eight-fold) degenerate for large positive (negative) $g$.

\section{two-leg ladder}
  \subsection{Phase diagram with nearest-neighbor hopping}

In this case the model can be converted into a particle number conserving Dirac model by defining:
\be c_{m}\equiv {\gamma_{m,0}+i(-1)^m\gamma_{m,1}\over 2}.\label{c2leg}\ee
We combine the Majoranas on each vertical link to make Dirac fermions. 
Thus
\bea \gamma_{m,0}&=&c_{m}+c_{m}^\dagger,\nonumber \\
\gamma_{m,1}&=&(-1)^mi(c_{m}^\dagger -c_{m}).\label{MD}
\eea
The horizontal hopping term becomes
\be H_0=2it\sum_m[c^\dagger_mc_{m+1}-c^\dagger_{m+1}c_m].\ee
The vertical hopping term vanishes with periodic boundary conditions in the $y$ direction since $\gamma_{m,0}\gamma_{m,1}+\gamma_{m,1}\gamma_{m,0}=0$. The interaction term becomes:
\be H_{int}=2g(2c^\dagger_mc_m-1)(2c^\dagger_{m+1}c_{m+1}-1).\ee
The factor of 2 in $H_{int}$ arises because there are 2 interaction terms for each $m$ due to the periodic boundary conditions in the $y$ direction; these 
both have the same sign. 
This is the standard spinless Dirac fermion model with nearest-neighbor interactions: attractive for $g>0$.  (We may eliminate the factor of $i$ from the 
hopping term by the transformation:
\be c_m\to i^mc_m,\label{tr}\ee
which shifts the wave-vector by $\pi /2$.  This transformation has no effect on the interaction term.) 
After a Jordan-Wigner transformation the Hamiltonian becomes:
\be H=\sum_m[t(\sigma^1_m\sigma^1_{m+1}+\sigma^2_m\sigma^2_{m+1})+2g\sigma^3_m\sigma^3_{m+1}],\label{eqxxz}\ee
the well-known XXZ model. For $|g|<1/2$, a Luttinger-liquid phase occurs. At $g=\pm 1/2$, transitions occur into ordered phases. 
Noting that
\be i\gamma_{m,0}\gamma_{m,1}=(-1)^m(2c^\dagger_mc_m-1)=(-1)^m\sigma^3_m,\label{FAF}\ee
we see that for $g>1/2$ the mean-field phase of Fig. \ref{fig:2d}(a) occurs and for $g<-1/2$ the mean-field phase of Fig. \ref{fig:2d}(b) occurs. 
[The transformation of Eq. (\ref{MD}) switches ferromagnetic with anti-ferromagnetic phases.]
It is interesting to note that, for the two-leg ladder, only the vertical order occurs, not the horizontal order. 
Furthermore, due to the absence of the vertical hopping term, filled or empty Dirac levels are degenerate, leading to a total of two ground states for 
 $g>0$ as well as $g<0$.  It is also interesting to note that at infinite $g/t$, corresponding to $t=0$, the case analyzed in [\onlinecite{Kamiya}], 
there is an infinite set of operators $\sigma^3_m$, commuting with the Hamiltonian. This corresponds to a trivial case of the  infinite number of ``intermediate symmetries'' of the 
compass model.  As expected in the 2D model \cite{Kamiya} at any finite temperature the broken symmetry is restored.

  \subsection{Effects of second-neighbor hopping}
  For the two-leg ladder, the effects of a second-neighbor hopping along the diagonals of the square plaquettes are easy to investigate.
 With periodic boundary conditions in the $y$ directions, the Hamiltonian has the additional term
 \be H_2=2it_2\sum_m[\gamma_{m,0}\gamma_{m+1,1}+\gamma_{m+1,0}\gamma_{m,1}].\ee
  In terms of the Dirac fermions \eqref{MD}, the above Hamiltonian becomes
 \be H_2=4t_2\sum_m (-1)^m(c_{m+1}c_m+{\rm H.c.}).\ee
The transformation of Eq. (\ref{tr}) gives
\be H_2=4t_2i\sum_m(c_{m+1}c_m-{\rm H.c.}).\ee
By making a Jordan-Wigner transformation, we then write this Hamiltonian in terms of the spin operators used in Eq.~\eqref{eqxxz} as follows
 \be H_2=-2t_2\sum_m(\sigma^1_m\sigma^2_{m+1}+\sigma^2_m\sigma^1_{m+1}).\ee
Making a $\pi/4$ rotation of all spin variables
 \[
\sigma^1_m\to (\sigma^1_m+\sigma^2_m)/\sqrt{2},\quad \sigma^2_m\to (\sigma^1_m-\sigma^2_m)/\sqrt{2}, 
 \]
  in the $1-2$ plane, we obtain:
 \be H=\sum_m[(t+2t_2)\sigma^1_m\sigma^1_{m+1}+(t-2t_2)\sigma^2_m\sigma^2_{m+1}+2g\sigma^3_m\sigma^3_{m+1}].\ee
 Note that while the $t_2$ term breaks time reversal in the 2D model it does not do it for the two-leg ladder with periodic boundary conditions due to the 
 absence of the vertical hopping term. We may define time reversal as
 \be \gamma_{m,n}\to (-1)^m\gamma_{m,n},\ \  i\to -i.\ee
 Choosing $t,t_2>0$, we get antiferromagnetic order in the 1 direction for sufficiently small $g$. This is a gapped phase. Although there is a spontaneously 
 broken symmetry in the spin model, with an order parameter of the form $ \sigma^1_m \cos \theta + \sigma^2_m \sin \theta$, this becomes a nonlocal 
 operator in the fermion model after the Jordan-Wigner transformation and there is no spontaneously broken symmetry in the fermion model. 
 This is also evident from the fact that, for $g=0$, it corresponds to a simple free fermion model with hopping and pairing terms. This model is equivalent to the Kitaev wire model \cite{Kitaev2001}, for which it is known that the topological degeneracy maps onto the symmetry breaking one under the Jordan-Wigner transformation.
 For sufficiently large positive $g$ the direction of the antiferromagnetic order should switch to the $3$ direction corresponding to 
 spontaneous breaking of the 
 $\sigma^3_m\to -\sigma^3_m$ symmetry and the type of Majorana order discussed above.

\section{four-leg ladder}
\subsection{Weak coupling}
We first discuss the noninteracting model for any even number of legs with PBC.  There are now $W$ values of $k_y$ for a ladder with $2W$ legs. 
We see from Eq. (\ref{E0}) that only the $k_y=0$ mode is gapless, with 
\be E_{\pm}=\pm 4t \sin k_x.\ee
This follows from Eq. (\ref{H0}) due to the absence of any even-odd coupling for $k_y=0$. So, the low energy theory contains only the $k_y=0$ mode. 
Fourier transforming with respect to $y$ only and keeping only $k_y=0$, we recover precisely the two-leg ladder model. Thus we see that, for any $W$
and small enough $|g|$, we recover the massless Luttinger liquid phase discussed above. However, we may expect that the transitions to 
gapped phases will occur at different values of $g$ and the universality classes of the phase transitions to gapped phases to be
 different than in the two-leg case.

\subsection{Strong coupling}
We now consider the four-leg ladder in the limit of $t=0$ for finite $g$; we later add an infinitesimal $t$. 
The Hamiltonian for $t=0$ can then be written as
\be H=-g\sum_m\sum_{j=0}^3(i\gamma_{m,j}\gamma_{m,j+1})(i\gamma_{m+1,j}\gamma_{m+1,j+1}).\label{Ht0}\ee
It turns out that in the limit of $t=0$, the fermion parity on each vertical rung is conserved. To see this, we combine pairs of Majoranas
 on vertical rungs into Dirac fermions as
\bea c_{m,1}&\equiv& (\gamma_{m,0}+i\gamma_{m,1})/2,\nonumber \\
c_{m,2}&\equiv& (\gamma_{m,2}+i\gamma_{m,3})/2.\eea
Inverting the above relations, we can then express the Majorana operators in terms of the Dirac operators as
\bea \gamma_{m,0}&=&c_{m,1}+c^\dagger_{m,1}, \quad \gamma_{m,1}=i(c_{m,1}^\dagger -c_{m,1}),\nonumber \\
\gamma_{m,2}&=&c_{m,2}+c^\dagger_{m,2},\quad \gamma_{m,3}=i(c_{m,2}^\dagger -c_{m,2}).\nonumber 
\eea
The terms in the Hamiltonian~\eqref{Ht0} contain products of four bilinears of the form $i\gamma_{m,j}\gamma_{m,j+1}$ for $j=0\dots 3$. Using the above relationships and taking the periodic boundary conditions into account, we then explicitly write all the above Majorana bilinears for $j=0\dots 3$ in terms of the Dirac fermions above:

\bea
i\gamma_{m,0}\gamma_{m,1}&=&2c_{m,1}^\dagger c_{m,1}-1,\label{gc} \\
i\gamma_{m,2}\gamma_{m,3}&=&2c_{m,2}^\dagger c_{m,2}-1,\nonumber \\
i\gamma_{m,1}\gamma_{m,2}&=&-c_{m,1}^\dagger c_{m,2}-c_{m,2}^\dagger c_{m,1}+c_{m,1}c_{m,2}+c^\dagger_{m,2}c^\dagger_{m,1},\nonumber \\
i\gamma_{m,3}\gamma_{m,0}&=&-c_{m,1}^\dagger c_{m,2}-c_{m,2}^\dagger c_{m,1}-c_{m,1}c_{m,2}-c^\dagger_{m,2}c^\dagger_{m,1}.\nonumber
\eea

From the equations~\eqref{gc}, we observe that $i\gamma_{m,j}\gamma_{m,j+1}$ for $j=0\dots 3$ either preserves the fermion number on the $m^{th}$ rung or changes it by $2$. Thus, the fermion
number is conserved mod 2 on each rung. Equivalently, fermion parity is conserved on each rung. We may simplify the $i\gamma_{m,j}\gamma_{m,j+1}$ 
operators depending on which parity sector we are in. Since for a fixed fermion parity, each rung forms a two-level system, it is convenient to identify the two states on each rung of given fermion parity with spin up and down 
states for a spin-1/2 variable, with 
\[\sigma^3_m|\uparrow\rangle_m=|\uparrow\rangle, \quad \sigma^3_m|\downarrow\rangle_m=-|\downarrow\rangle_m.\]
Suppressing the $m$ index for brevity, we summarize the identification of the up and down spins below
{\renewcommand{\arraystretch}{1.4}
\begin{center}
\begin{tabular}{c|c}
\hline 
\hline 
{\it Even fermion parity} & {\it Odd fermion parity}\\ 
\hline 
 $|\downarrow \rangle \equiv |0\rangle,\quad |\uparrow \rangle \equiv c_1^\dagger c_2^\dagger |0\rangle$ &  $|\downarrow \rangle \equiv c_1^\dagger|0\rangle,\quad |\uparrow \rangle \equiv  c_2^\dagger |0\rangle$ \\ 
\hline 
\hline 
\end{tabular} 
\end{center}
Using the above spin notation, we then identify the Majorana bilinears of Eq.~\eqref{gc} with spin operator given by Pauli matrices. The results are summarized in Table~\ref{tab}.
\begin{center}
\begin{table}[b]
\begin{tabular}{c|c}
\hline 
\hline 
{\it Even fermion parity} & {\it Odd fermion parity}\\ 
\hline 
 $i\gamma_{m,0}\gamma_{m,1}=\sigma^3_m$ &  $i\gamma_{m,0}\gamma_{m,1}=-\sigma^3_m$ \\ 
  $i\gamma_{m,2}\gamma_{m,3}=\sigma^3_m$ &  $i\gamma_{m,2}\gamma_{m,3}=\sigma^3_m$ \\ 
    $i\gamma_{m,1}\gamma_{m,2}=\sigma^1_m$ &  $i\gamma_{m,1}\gamma_{m,2}=-\sigma^1_m$ \\ 
      $i\gamma_{m,3}\gamma_{m,0}=-\sigma^1_m$ &  $i\gamma_{m,3}\gamma_{m,0}=-\sigma^1_m$ \\
\hline 
\hline 
\end{tabular}\caption{The strong coupling mapping between spin variables and Majorana bilinears.\label{tab}}
\end{table} 
\end{center}
The derivation of the above is straightforward. As an example consider $i\gamma_{m,1}\gamma_{m,2}$ in the odd fermion parity sector. Using Eq.~\eqref{gc}, we can show that 
\be
i\gamma_{1}\gamma_{2}c^\dagger_1|0\rangle=-c^\dagger_2|0\rangle,\quad
i\gamma_{1}\gamma_{2}c^\dagger_2|0\rangle=-c^\dagger_1|0\rangle,\nonumber
\ee
In other words, this operator is equivalent to a spin flip and a sign change and can be identified with $-\sigma^1$.

%
%
%

Assuming the same fermion parity on every rung, we see from the table above that for both parities, the Hamiltonian~\eqref{Ht0} maps to the following spin model:
\be H=-2g\sum_m(\sigma^3_m\sigma^3_{m+1}+\sigma^1_m\sigma^1_{m+1}),\ee
where the factor of 2 come from adding the contributions of $i\gamma_{m,0}\gamma_{m,1}$ and $i\gamma_{m,2}\gamma_{m,3}$ for the $\sigma^3_m\sigma^3_{m+1}$ term, and $i\gamma_{m,1}\gamma_{m,2}$ and $i\gamma_{m,3}\gamma_{m,0}$ for the $\sigma^1_m\sigma^1_{m+1}$ term.
If any neighboring pair of rungs, say for $m$ and $m+1$, have opposite fermion parity, we see from Eq.~\eqref{Ht0} that the contributions of $i\gamma_{m,0}\gamma_{m,1}$ and $i\gamma_{m,2}\gamma_{m,3}$ have opposite signs and cancel out (and similarly for $i\gamma_{m,1}\gamma_{m,2}$ and $i\gamma_{m,3}\gamma_{m,0}$ for the $\sigma^1_m\sigma^1_{m+1}$). Thus for neighboring rungs with opposite fermion parity, all the terms connecting the rungs in the Hamiltonian vanish.

Two neighboring rungs with opposite fermion parity remove a negative contribution to the ground-state energy.
Therefore the ground state must have the same fermion parity on every rung. This spin model has no broken symmetries and is gapless. However, there is actually a two-fold 
ground state degeneracy--- even or odd fermion parity.  Again there is an infinite number of conserved quantities, the fermion parity on every rung, as in the compass model, relevant to the 2D case~\cite{Kamiya}. 
Unlike the two-leg case, the four-leg model is massless at $t=0$. This massless behavior is analogous to that of the compass model. 

Now consider turning on an infinitesimal hopping term. The hopping terms in the vertical direction on the $m^{th}$ rung can be written by adding the contributions of all the terms in the table above:
\bea it\sum_j \gamma_{m,j}\gamma_{m,j+1}&=&2t\sigma^3_m  \ \  (\hbox{even fermion parity}),\nonumber \\
 it\sum_j \gamma_{m,j}\gamma_{m,j+1}&=&-2t\sigma^1_m  \nonumber\ \  (\hbox{odd fermion parity}).\eea

 On the other hand, hopping terms in the horizontal direction, i.e., $it\gamma_{m,j}\gamma_{m+1,j}$, contain terms with a single creation or annihilation operator for each rung. Therefore, they change the fermion number on each rung by $\pm 1$, reversing the fermion parity. consider these terms as a perturbation around the ground state of the $t=0$ model, for which all rungs have the same fermion parity. Upon acting on the ground state, we get a state 
 in which two of the neighboring bonds have opposite fermion parity, raising the energy by order $O(g)$. Specifically, acting with the hopping term on the $m\leftrightarrow (m+1)$ bond gives opposite fermion parity on the 
 $(m-1)\leftrightarrow m$ and $(m+1)\leftrightarrow (m+2)$ bonds.  Such high-energy states contribute at second-order in perturbation theory and have effects of order $O(t^2/g)$. Thus, the leading-order Hamiltonian to first-order in the small hopping $t$ can be written for the two cases of even and odd fermion parity on every rung as
 \bea H_{\rm even}&=&\sum_m[-2g(\sigma^3_m\sigma^3_{m+1}+\sigma^1_m\sigma^1_{m+1})+2t\sigma^3_m], \nonumber \\
 H_{\rm odd} &=&\sum_m[-2g(\sigma^3_m\sigma^3_{m+1}+\sigma^1_m\sigma^1_{m+1})-2t\sigma^1_m].\nonumber \eea
The behavior of the above Hamiltonians are easy to understand. For $t=0$, we have an XY chain in the 1-3 plane. Furthermore, the sign of $t$ is unimportant as it can be changed by a $\pi$ rotation. The two Hamiltonians for the even and odd sector are also equivalent as they are related by a $\pi/2$ rotation.
In our convention, with $t>0$, the even (odd) case has an in-plane magnetic field in the $+3$ ($-1$) direction.

Let us first focus on the $g>0$ case, where the XY interaction is ferromagnetic. The spins have a tendency to align in the plane and any nonzero filed in the plane explicitly breaks the symmetry and picks a direction. Assuming $t>0$ (without loss of generality), for $g>0$, $\langle \sigma^3_m\rangle <0$ for even fermion parity and $\langle \sigma^1_m\rangle >0$ for odd fermion parity, since the hopping term give a filed in the positive 3 (negative 1) for even (odd) parity.

The interpretation of these states in terms of the Majorana fermions can be obtained from the table above. For instance, the state with $\langle \sigma^3_m\rangle <0$ for even parity has $\langle i\gamma_{m,0}\gamma_{m,1}\rangle=\langle i\gamma_{m,2}\gamma_{m,3}\rangle<0$, which corresponds to the two Dirac fermion levels formed on vertical bonds $0-1$ and 
$2-3$ being empty. For odd fermion parity, we see that  $\langle \sigma^1_m\rangle >0$ corresponds to $\langle i\gamma_{m,1}\gamma_{m,2}\rangle=\langle i\gamma_{m,3}\gamma_{m,0}\rangle<0$, i.e., the two Dirac fermion levels formed 
 on vertical bonds $1-2$ and $3-0$ being empty. These correspond precisely to the two 
 mean-field 2D ground states of Fig. (\ref{fig:2d}).  The ladder geometry breaks the $\pi/2$ rotation symmetry of the 2D square lattice and favors Dirac fermions forming on vertical, not horizontal links. So, the 
 number of ground states is 2, not 4.

It is also interesting to note that, since the expectation value of $\sigma^1$ vanishes with respect to an eigenstate of $\sigma^3$, $\langle \sigma^1_m\rangle$ is presumably zero in the even fermion parity sector, for which $\sigma^3$ is condensed with $\langle \sigma^3_m\rangle <0$. Thus 
 $\langle i\gamma_{m,1}\gamma_{m,2}\rangle =\langle i\gamma_{m,3}\gamma_{m,0}\rangle =0$. This is similar to the mean-field picture in the 2D case, where $\gamma_{m,0}$ and 
 $\gamma_{m,1}$ combine to form a Dirac fermion and also $\gamma_{m,2}$ and $\gamma_{m,3}$. With these pairs of Majoranas combining, two Majoranas from different pairs, e.g., $\gamma_{m,1}$ and $\gamma_{m,2}$ remain unentangled so their product has zero expectation value (and similarly for the $\gamma_{m,3}\gamma_{m,0}$ product). 
The same argument in the odd fermion parity sector leads to
 $\langle \sigma^3\rangle=0 $ corresponding to $\gamma_0-\gamma_1$ and $\gamma_2-\gamma_3$ not being entangled. 
 
We now consider the $g<0$ case, where the XY chain is antiferromagnetic. The effect of the $t$ term is again a  uniform in-plane magnetic field. In this case, we expect the direction of the in-plane antiferromagnetic order to be perpendicular to the field. It is well known that classically, minimizing the energy $\sum_m [J\cos(\theta_m-\theta_{m+1})-2h\cos(\theta_j)]$ of antiferromagnetically interacting XY spins in an in-plane field leads to canting in the direction of the field and antiferromagnetic ordering perpendicular to the field. We expect the same physics to appear quantum mechanically. Indeed, a variational quantum wave function made of a product of in-plane spin-$1\over 2$ variables $\cos(\theta_m/2)|\uparrow\rangle_m+\sin(\theta_m/2)|\downarrow\rangle_m$ gives the same energy expectation value as the above classical energy. 
Importantly, there is no violation of the Mermin-Wagner theorem since there is no U(1) continuous symmetry once the magnetic field is present. There is a unique classical direction for the antiferromagnetic order (up to a sign) and breaking of the discrete $Z_2$ symmetry in the quantum model is expected. These states are shown in Fig.~\ref{fig:demo}.
\begin{figure}
\epsfxsize=8 cm
\includegraphics*[width=\columnwidth]{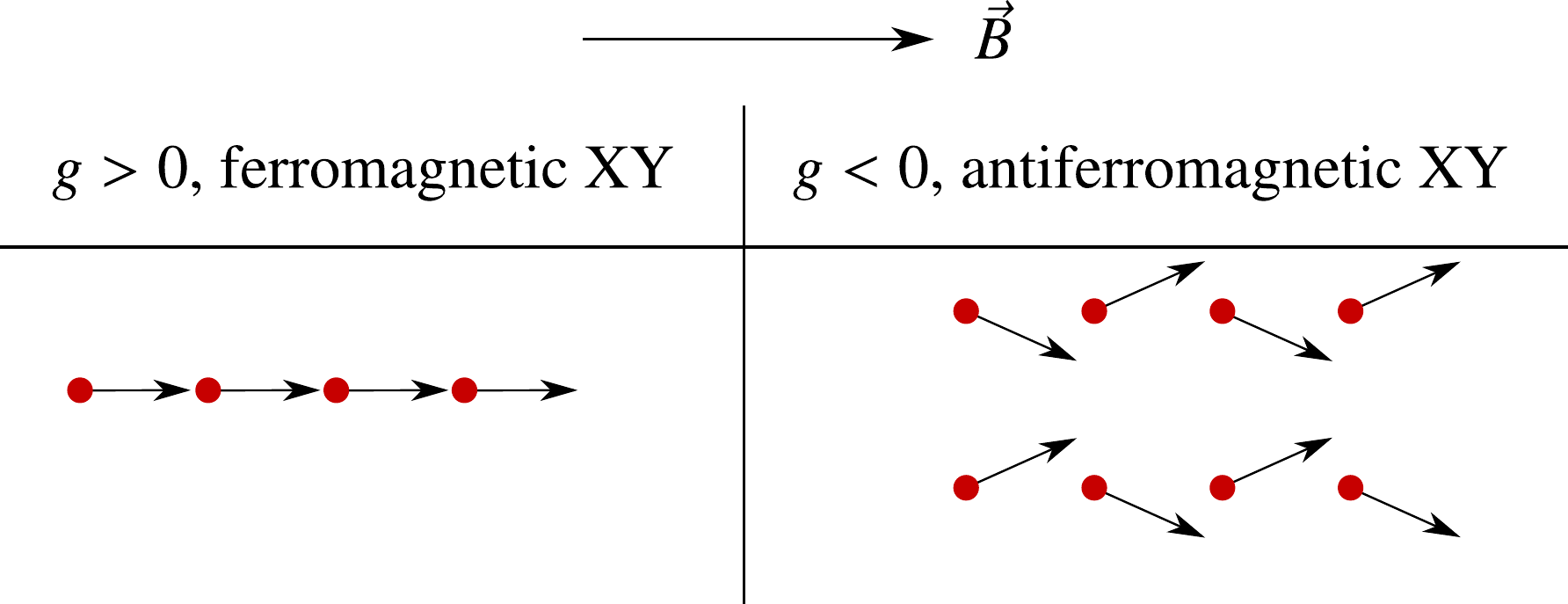}
\caption{A cartoon picture for the ground states of the ferromagnetic and antiferromagnetic XY chain in an in-plane field $\vec B$.}
\label{fig:demo}
\end{figure}

 Thus, because the even (odd) sector has a field in the 3 (1) direction, we expect antiferromagnetic order to appear in the perpendicular, i.e., 1 (3) direction:
 \bea \langle \sigma^1_m\rangle &\propto& (-1)^m \ \  \hbox{(even fermion parity)}\nonumber \\
 \langle \sigma^3_m\rangle &\propto& (-1)^m \ \  \hbox{(odd fermion parity)}.\eea
Again, we see that these states correspond to forming Dirac levels on vertical bonds. In the even parity case, we have $\langle i\gamma_{m,1}\gamma_{m,2}\rangle=-\langle i\gamma_{m,3}\gamma_{m,0}\rangle \propto(-1)^m$. In the odd parity case, we have $-\langle i\gamma_{m,0}\gamma_{m,1}\rangle=\langle i\gamma_{m,2}\gamma_{m,3}\rangle \propto(-1)^m$. Thus, the levels for both parity sectors are alternating filled and empty along each row.  However, unlike the mean-field prediction of the 2D square lattice, shown in Fig. ({\ref{fig:2d}), the two levels
 on each rung have opposite filling, as illustrated in Fig. (\ref{fig:4legVAF}). For the odd parity case, this result is very simple to understand: the total number of electrons on each rung is 1, so if the $0-1$ level has filling 
$<1/2$ then the $2-3$ level must have filling $>1/2$ and vice versa.

%

  \begin{figure}
\epsfxsize=8 cm
\includegraphics*[width=.8\columnwidth]{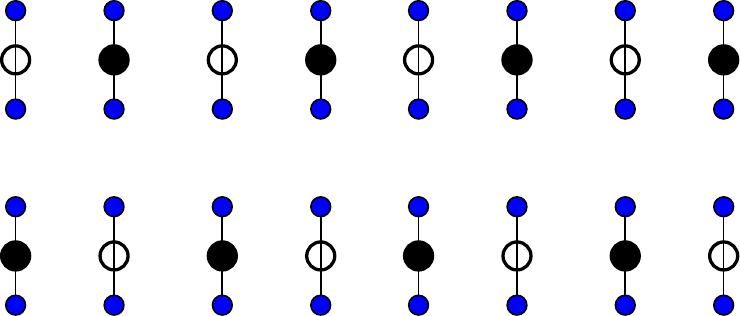}
\caption{1 of the 4 ground states occurring in the 4-leg ladder for $g\ll -t$.}
\label{fig:4legVAF}
\end{figure}

We now consider the nature of the phase transition that occurs at $t=0$, into a gapped phase. Since the Hamiltonians for the two sectors are equivalent, we use the even sector notation. It is convenient to rotate the spins $(\sigma_m^3,\sigma_m^1)\to (\sigma_m^1,\sigma_m^2)$ to obtain a more standard model $H=\sum_m[-2g(\sigma^1_m\sigma^1_{m+1}+\sigma^2_m\sigma^2_{m+1})+2t\sigma^1_m]$. For $g>0$ we perform another $\pi$ rotation on every other site to change the sign of $g$ at the expense of staggering the in-plane field. We can then write
 \bea H &=&\sum_m[2|g|(\sigma^1_m\sigma^1_{m+1}+\sigma^2_m\sigma^2_{m+1})+2t(-1)^m\sigma^1_m]\ \  (g>0)\nonumber \\
  &=&\sum_m[2|g|(\sigma^1_m\sigma^1_{m+1}+\sigma^2_m\sigma^2_{m+1})+2t\sigma^1_m]\ \  (g<0).
 \eea
 In both cases we obtain the antiferromagnetic XY model with an in-plane staggered field for $g>0$ and an in-plane uniform field for $g<0$.
 
 A field theoretical analysis based on the Abelian bosonization is helpful in understanding the behavior of the above model. Let us neglect the in-plane field for now. The XY term simply maps to Dirac fermions hopping on the lattice. It is well known that this theory can be bosonized and gives a Luttinger liquid:
 \be 
 H\big |_{t=0}={v\over 2}\int dx\left\{K[\partial_x \phi (x)]^2+{1\over K}[\partial_x \theta (x)]^2\right\},
 \ee
 where $K$ is the Luttinger parameter and $v$ is the velocity. The bosonic fields $\theta$ and $\phi$ satisfy the commutations relations
 \[
 [\phi(x),\theta(x')]=i \Theta(x-x'),
 \] 
where $\Theta$ is the Heaviside step function.
 The presence of a $\sigma^3_m\sigma^3_{m+1}$ interaction would change the Luttinger parameter from the noninteracting value $K=1$, but here we do not have this interaction and therefore $K=1$. 
 
The above bosonic fields can be written, in turn, in terms of right- and left-movers, which are helpful for calculating the scaling dimensions of various operators: 
 \be \phi=\phi_L+\phi_R,\ \  \theta =\phi_L-\phi_R ,\ee
where
\[
\langle e^{i2\sqrt{\pi}\phi_{L,R}(x)} e^{-i2\sqrt{\pi}\phi_{L,R}(0)}\rangle\sim 1/x.
\]
The bosonized expression for the spin operator is~\cite{Giamarchi}
 \be \sigma^1_m= \cos (\sqrt{\pi }\theta )[(-1)^mA+B\cos \sqrt{4\pi}\phi ].\ee 
 Expanding out the equal-time correlation function $\langle \sigma^1_0\sigma^1_m\rangle$, we find terms of the following types:
\bea
(-1)^m \langle e^{i\sqrt{\pi}[\phi_{L}(m)-\phi_{R}(m)]} e^{-i\sqrt{\pi}[\phi_{L}(0)-\phi_{R}(0)]}\rangle&\sim& {(-1)^m\over m^{1/2} },\nonumber\\
\langle e^{i\sqrt{\pi}[3\phi_{L}(m)+\phi_{R}(m)]} e^{-i\sqrt{\pi}[3\phi_{L}(0)+\phi_{R}(0)]}\rangle&\sim& {1\over m^{5/2} },\nonumber \\
\langle e^{i\sqrt{\pi}[\phi_{L}(m)+3\phi_{R}(m)]} e^{-i\sqrt{\pi}[\phi_{L}(0)+3\phi_{R}(0)]}\rangle&\sim& {1\over m^{5/2} }.\nonumber 
\eea

We first consider the $g>0$ case, for which the leading nonoscillatory term in the equal-time correlator of the staggered field $(-1)^m\sigma^1_m$ goes as $1/m^{1/2}$. This implies that this perturbation has a scaling dimension $d=1/4$. We use RG to determine the effect of this perturbation. In (1+1) dimensions, the action changes by an integral of the perturbation operator over the two-dimensional space-time. Thus, upon an RG transformation $x\to\alpha x$, the coupling constant $t$ transforms as $t\to \alpha^{2-d}t$. Therefore, for $d>2$ ($d<2$), the perturbation becomes less (more) important at larger length scales, i.e., it is irrelevant (relevant) in the RG sense. A relevant operator of dimension $d$ is expected to break the conformal symmetry and open a mass gap scaling as
 \be m\propto t^{1/(2-d)}=t^{4/7}.\ee
  Including $g$, on dimensional grounds, the gap scales as $|t|^{4/7}g^{3/7}$ for $g\to +\infty $.
 
 We now consider the $g<0$ case, where the perturbation is a uniform field $\sigma^1_m$. The leading nonoscillatory term in the equal-time correlator of $\sigma^1_m$ goes as $1/m^{5/2}$. Thus, the uniform field has dimension $d=5/4<2$ and is relevant. We might expect it to produce a mass scaling with exponent $1/(2-d)=4/3$. 
 However, this argument is inconsistent with the expectation of a gapped phase with antiferromagnetic order in the $2$ direction. If such order indeed emerges, we see from the bosonized expression
 \be \sigma^2_m= \sin (\sqrt{\pi }\theta )[(-1)^mA+B\cos \sqrt{4\pi}\phi ].\ee 
 that the field $\phi$ must be pinned. However, if $\phi$ is pinned, $\theta$ 
 fluctuates strongly so this uniform field term becomes irrelevant.  However, at second order in $t$, we should generate a relevant $\cos \sqrt{4\pi}\theta$ term of dimension $d=1$, which can pin $\phi$.  In fact, this corresponds to anisotropic exchange, $J^y>J^x$ (with a $J^x\sigma^1_m\sigma^1_{m+1}+J^y\sigma^2_m\sigma^2_{m+1}$ coupling), also favoring antiferromagnetic order. Actually the XY model with $J^x\neq J^y$ is 
 exactly solvable by the Jordan-Wigner transformation and certainly has a gap. Once we add a uniform field there is no symmetry forbidding anisotropic exchange so it looks reasonable for the anisotropy to be generated in RG.
This suggests that the relevant coupling constant is $\propto t^2$ and of dimension 1, leading to 
 a gap $\propto t^2$, rather than $|t|^{4/3}$.  Including $g$, the gap thus scales as $t^2/|g|$ for $g\to -\infty$. 
 
 We were able to verify the above predictions numerically for $g/t\to \pm\infty$ by computing the energy gap $\Delta E$ (in the even fermion parity sector) with antiperiodic boundary conditions on the majorana fermions in the $x$ direction, with DMRG for various system sizes, and extrapolating them to the thermodynamic limit. Let us first discuss the $g>0$ case, where the extrapolation was very straightforward. We fixed $g=1$ and used several small values of $t=0.01\dots 0.05$ for system sizes $N=40,50\dots 80$. the numerically computed gaps for finite systems are extrapolated to the thermodynamic limit through a linear fit to $1/N$. Plotting the extrapolated values $\ln (\Delta E)$ as a function of $\ln(t)$, as shown in Fig. \ref{fig:gg}, we find a very good linear fit with slope $0.58$ that is very close to $4/7\approx 0.57$. 
 \begin{figure}
\epsfxsize=8 cm
\includegraphics*[width=0.9\columnwidth]{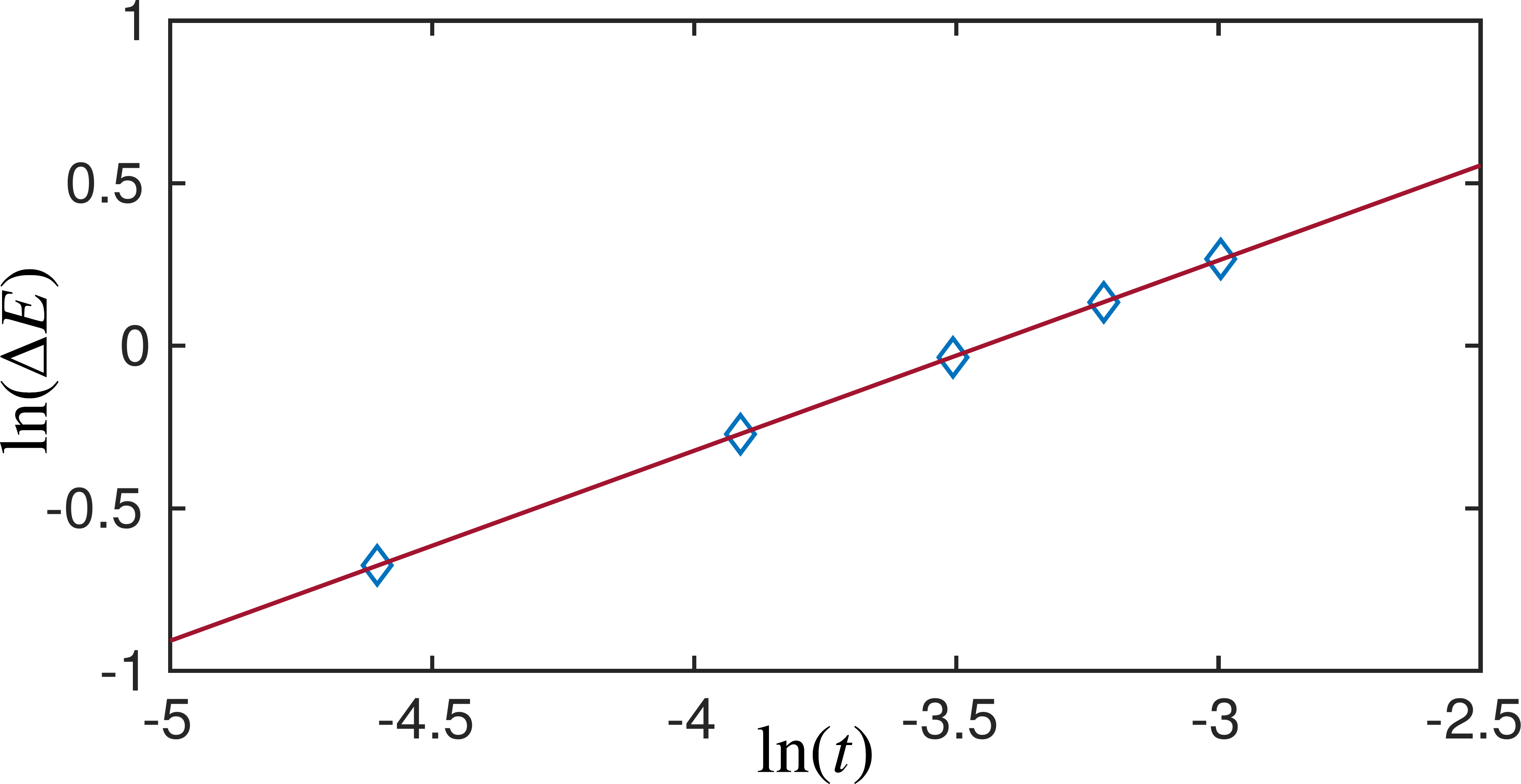}
\caption{A linear fit of $\ln(\Delta E)$ as a function of $\ln (t)$ for $g=1$ and small $t$ supports a $t^{4/7}$ scaling of the gap.}
\label{fig:gg}
\end{figure}

For $g<0$, the extrapolation is more subtle. As shown in Fig.~\ref{fig:gg2}(a), the numerically computed gaps for small $t$ (calculated by setting $g=-1$) fit very well to a linear function of $1/L$. Interestingly, they all appear to extrapolate to \textit{negative} values. Therefore, it seems that the numerically computed ground (first excited) state for finite systems actually extrapolates to the first excited (ground) state in the thermodynamic limit, for very small $t$. However, the system sizes for which the crossing would occur are very large for extremely small $t$, and it becomes difficult to numerically observe the crossing directly. Note that, as shown in Fig.~\ref{fig:schemFD}, the extrapolated value of $g_2\approx -21.1$ (using the linear fit shown in Fig.~\ref{fig:extrap}) suggests $|g_3|>|g_2|\approx 21.1$, so we expect this extrapolation to break down at least for $t>0.05$ (maybe for any infinitesimal $t$), although it is known to capture the correct behavior for $t=0$ (the XY point). Indeed, for larger $t$, where we are in gapless phases from entanglement entropy calculations, small-system extrapolation may suggest a similar crossing. However, it can be seen that a new pattern emerges in the energy gaps when the finite size $\Delta E$ approaches zero, with $\Delta E $ beginning to increase (presumably extrapolating to zero after oscillations). Consequently, the linear extrapolation must be taken with a grain of salt.

The finite size effects are much larger in this region of the phase diagram, and it is very hard to determine the value of $g_3$ due to a potential change in the linear dependence of $\Delta E$ on $1/L$.  Nevertheless, if we take the linear extrapolation at face value and assume that the first excited state indeed extrapolates to the ground state in a very small region of the phase diagram (where a gap is expected on theoretical grounds), the extrapolated gaps would be consistent with a $t^2$ scaling as seen in Fig.\ref{fig:gg2}(b). 
 \begin{figure}
\includegraphics*[width=\columnwidth]{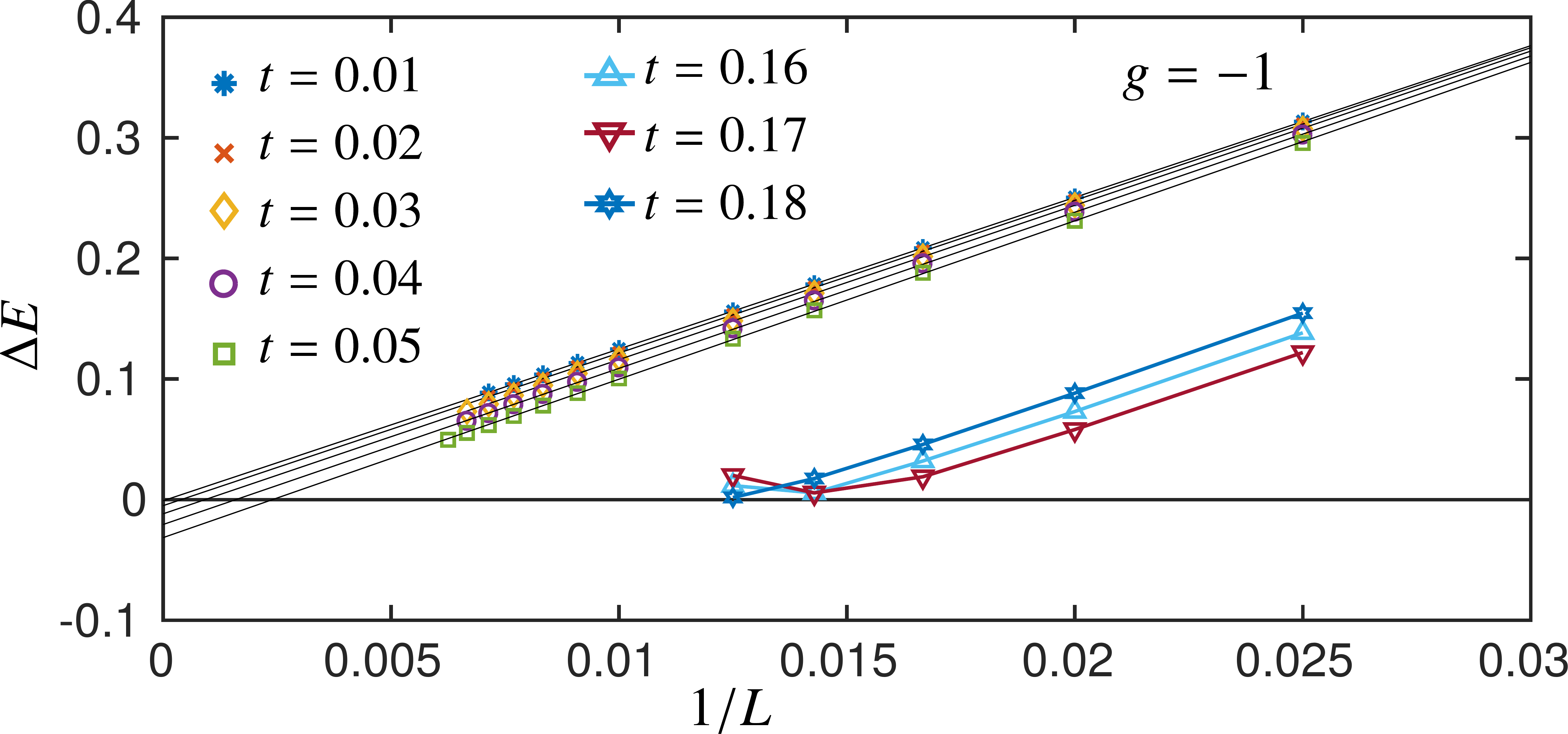}
\caption{The extrapolation of the energy gap for large negative $g$. For very small values of $t$, we observe a linear fit, which suggests an extrapolation of $\Delta  E$ to negative values, i.e., a crossing of the finite-system ground and first-excited states. However, this extrapolation scheme must break down either immediately of for a critical value of $t$. For larger $t$ (in other phases), small system extrapolation suggests a similar linear fit but the behavior changes when the finite-system gaps approach zero.
\label{fig:gg1}}
\end{figure}

 \begin{figure}
\includegraphics*[width=0.8\columnwidth]{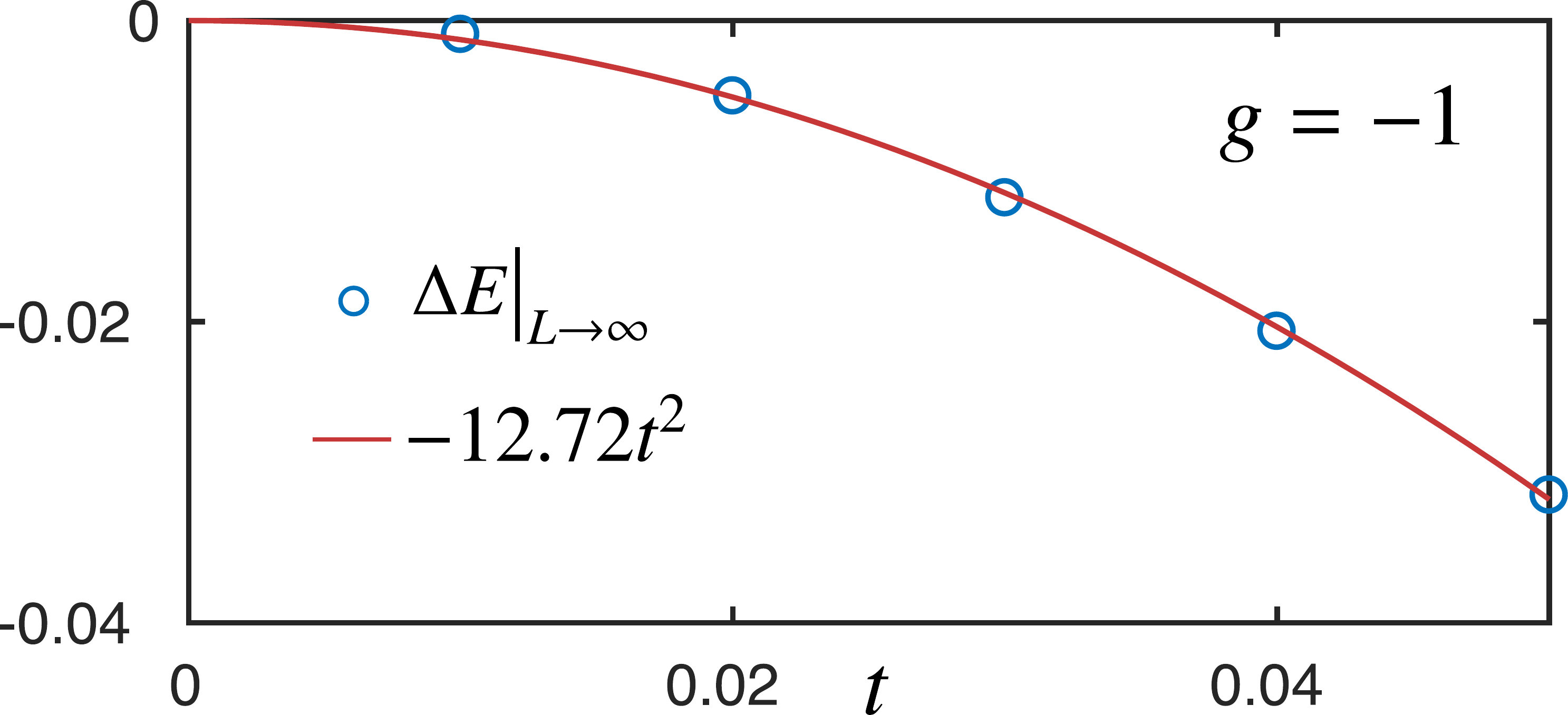}
\caption{While we have evidence that the linear extrapolation breaks down at least for $t$ larger than a finite value, we know that it is correct for $t=0$. This suggests that there may be a small range of $t$, corresponding to the gapped strong-coupling case for $g<0$, where the finite-system levels cross upon increasing system size and the linear extrapolation works. Assuming the survival of the linear extrapolation in the gapped phase, the extrapolated gaps indeed scale as $t^2$, as predicted.
\label{fig:gg2}}
\end{figure}
 
 \section{Open boundary conditions in $y$-direction}
Throughout this paper, we primarily focus on periodic boundary conditions in the $y$ direction. However, in this section we comment on some of the features of open boundary conditions. Open boundary conditions eliminate the massless $k_y=0$ mode in the noninteracting model.  In the limit  $W\to \infty$, the 2D model, we expect this to be unimportant. However, for finite-width ladders, it has a large effect. 
 
 In the two-leg case, open boundary conditions lead to a vertical hopping term (recall that it canceled out with periodic boundary conditions). We see from Eq. (\ref{FAF}) that this changes the spin Hamiltonian to 
 \be H=\sum_m[t(\sigma^1_m\sigma^1_{m+1}+\sigma^2_m\sigma^2_{m+1})+g\sigma^3_m\sigma^3_{m+1}+t(-1)^m\sigma^3_m],\ee
 adding a staggered field in the $3$ direction, which can produce a gap, even for $g=0$.  This gap is present for all $g\geq 0$. However, for $g<0$, where 
 the exchange term is ferromagnetic this staggered field leads to frustration.  At large enough $g<0$, a ferromagnetic order occurs. There is 
 an intermediate range of $g<0$ where a massless Luttinger liquid phase occurs.\cite{Alcarez,Okamoto}
 
 In the four-leg case, the gapless phase at small $g$ is again eliminated.  The gapless phase at $t=0$ is also eliminated. This follows because the plaquette 
 interaction involving rows 3 and 0 does not appear so, from Table \ref{tab}, the Hamiltonian becomes 
 \be H=-g\sum_m[2\sigma^3_m\sigma^3_{m+1}+\sigma^1_m\sigma^1_{m+1}].\ee
 This model should be gapped with order in the $3$ direction, ferromagnetic for $g>0$ and antiferromagnetic for $g<0$. 
 Again this corresponds to Dirac fermion formation on vertical bonds but now only occurring on $0-1$ and $2-3$ bonds. 
A small hopping term adds a perturbation, which an be written in terms of the spin variable from the table~\ref{tab}:
\bea \delta H&=&t\sum_m(2\sigma^3_m+\sigma^1_m), \quad {\rm (even~fermion ~parity)},\nonumber \\
 \delta H&=&t\sum_m(-\sigma^1_m), \quad {\rm (odd ~fermion ~parity)},\nonumber 
\eea
For $g>0$, we have ferromagnetic order in the 3 direction. The $\sigma^1$ field in the perpendicular direction has no first-order effect on the energy so in the odd fermion-parity case, the energy remains unchanged. In the even fermion-parity case, the $\sigma^3$ field directly couples to the magnetic order parameter, selecting one of the ferromagnetic states and reducing its energy. Therefore, we get a unique ground state with the Dirac levels empty. For $g<0$, we have an antiferromagnetic state in the 3 direction and neither the $\sigma^1$ nor the $\sigma^3$ fields  change the two ground state energy to first order in $t$.

 \section{Kosterlitz-Thouless Transition}
 Here we argue that the first transition upon increasing $g>0$ from the noninteracting point, at $g=g_{KT}\approx 0.8$, as we will see in the numerical studies of Sec.~\ref{sec:num}, is of the Kosterlitz-Thouless type. We use the mapping of the two $k_y$-momentum channels to a charged fermion ladder as discussed in the Appendix. We first consider the sum of the noninteracting Hamiltonian in Eq. (\ref{eq:tran}) and  the interacting Hamiltonian 
 in (\ref{Hintc}), taking into account only the intrachannel interactions. Then, at least for sufficiently small $g/t$, we see that the $k=0$ sector is in a gapless Luttinger liquid phase, the standard spinless fermion model with nearest-neighbor 
 interactions,  while the $k=\pi$ sector is gapped.  Now let's include the interchannel interactions and integrate out the gapped $k=\pi$ sector. To first order in $g$, we can just replace 
 $(c^\dagger_{m+1,\pi}+c_{m+1,\pi})(c^\dagger_{m,\pi}-c_{m,\pi})$ and $(c^\dagger_{m,\pi}+c_{m,\pi})(c^\dagger_{m+1,\pi}-c_{m+1,\pi})$ by their expectation values. Since, ignoring interchannel interactions, the Hamiltonian preserves 
 charge, these reduce to $- (c^\dagger_{m,\pi}c_{m+1,\pi}+c^\dagger_{m+1,\pi}c_{m,\pi})$.  Next, we note that the expectation value of $(c^\dagger_{m,\pi}c_{m+1,\pi}+c^\dagger_{m+1,\pi}c_{m,\pi})$ is independent of $m$. 
 This follows, despite the breaking of translation symmetry by the $(-1)^m(2c^\dagger_{m,\pi}c_{m,\pi}-1)$ term in Eq. (\ref{eq:tran}) due to the parity symmetry, $m\to -m$. 
  This term simply renormalizes the hopping term in the $k=0$ sector 
 and does not break the U(1) symmetry. At next order in $g$, we generate interactions of the form:
\bea
\delta H&=&\sum_{m,m'}[\lambda_{m-m'}(c^\dagger_{m,0}+c_{m,0})(c^\dagger_{m+1,0}-c_{m+1,0})\nonumber \\
  &\cdot& (c^\dagger_{m'+1,0}+c_{m'+1,0})(c^\dagger_{m',0}-c_{m',0})\nonumber \\
  &+&g_{m-m'}(c^\dagger_{m,0}+c_{m,0})(c^\dagger_{m+1,0}-c_{m+1,0})\nonumber \\
 &\cdot& (c^\dagger_{m',0}+c_{m',0})(c^\dagger_{m'+1,0}-c_{'m+1,0})\nonumber \\
  &+&g_{m-m'}(c^\dagger_{m+1,0}+c_{m+1,0})(c^\dagger_{m,0}-c_{m,0})\nonumber \\
 &\cdot&  (c^\dagger_{m'+1,0}+c_{m'+1,0})(c^\dagger_{m',0}-c_{m',0})
 ]
\eea
where the $g_{m-m'}$ and $\lambda_{m-m'}$ couplings drop off exponentially with distance and are even functions of
$m-m'$. These do break the U(1) symmetry. It can be checked that terms of charge $\pm 2$ cancel, so the U(1) breaking
part is
$\sum_{m,m'}(2g_{m-m'}-\lambda_{m-m'})c^\dagger_{m,0}c^\dagger_{m+1,0}c^\dagger_{m',0}c^\dagger_{m'+1,0}+{\rm H.c.}$
 To study its effect we can bosonize the fermion model. The only potentially relevant term comes from
 \be \psi^\dagger_R\partial_x\psi^\dagger_R\psi^\dagger_L\partial_x\psi^\dagger_L+h.c.\propto \cos [4\sqrt{\pi /K}\theta].
 \label{bos}\ee
 This has dimension $d=4/K$.  $K=1$ for $g=0$ and $K$ decreases for $g>0$ corresponding to effectively repulsive
 interactions, as we see from Eq. (\ref{Hintc}).  Thus, this symmetry breaking operator is irrelevant for $g>0$.
 On the other hand, the symmetry preserving interactions lead to the standard umklapp term
 \be \psi^\dagger_L\partial_x\psi^\dagger_L\psi_R\partial_x\psi_R+h.c.\propto \cos (4\sqrt{\pi K}\phi),\ee
 of dimension $d=4K$. This becomes relevant for sufficiently strong $g$, at $K=1/2$, where a charge-density-wave transition
 occurs, corresponding to the usual Kosterlitz-Thouless transition in XXZ spin chain at the Heisenberg point. 
 
 The left-moving single fermion operator bosonizes as
 \be \psi_L\propto e^{i\sqrt{4\pi}\phi_L}\ee
 at $g=0$ where $K=1$.  Writing
 \be \phi=\phi_L+\phi_R,\ \  \theta =(\phi_L-\phi_R)/K ,\ee
 this becomes
 \be \psi_L\propto e^{i\sqrt{\pi}(\phi -\theta )}=e^{i\sqrt{\pi}[(1+1/K)\phi_L+(1-1/K)\phi_R]/2}\ee
 of dimension
 \be d=(1/4)(K+1/K).\ee
 This gives $d=1/2$ for free fermions, $K=1$ and $d=5/8$ at the KT point, $K=1/2$. The equal time Green's function for the  fermion decays with exponent $2d$: $1$ for free fermions and $5/4$ at the KT point. 
 The central charge is $c=1$ along the entire critical line including at the KT point. 
 
For $g>g_{KT}$ we expect a charge density wave.  The order parameter is
\be(-1)^m(2c^\dagger_{m,0}c_{m,0}-1).\ee
However, when relating the above expression to the original Majorana fermions, we have to take into account the transformation~\eqref{eq:tran}, which eliminates the factor of $(-1)^m$, when going from the new to the old Dirac fermions so the order parameter reduces to $i\gamma^e_{m,0}\gamma^o_{m,0}$ in terms of the original Majoranas. We can thus write the order parameter as
\be(-1)^m\langle \sigma^z_{m,0}\rangle  ={i\over 2}\sum_{n=0}^3(-1)^n\langle \gamma_{m,n}\gamma_{m,n+1}\rangle.\ee
%
%
%
 The expression above corresponds to the order of Fig. (\ref{fig:4legVAF}).
  Conversely
 \be (-1)^m\langle \sigma^z_{m,\pi}\rangle ={i/2}\sum_{n=0}^3\langle \gamma_{m,n}\gamma_{m,n+1}\rangle \ee
 as we see from Eq. (\ref{VHT}), which does not break any symmetries. 
\section{Numerical phase diagram of the four-leg ladder}\label{sec:num}
As discussed above, we expect gapped phases with certain symmetry breaking patterns at the two strongly interacting regimes $g/t\to\pm \infty$. The noninteracting point is a critical phase with central charge $c=1$, described by a noninteracting Luttinger liquid. Motivated by the results of the single chain, we expect transitions between possibly several critical phases before reaching the broken-symmetry gapped phases. On the positive $g$ side, however, our theoretical predictions support a single Kosterlitz-Thouless transition from the Luttinger liquid phase. In the chain, we had one (supersymmetric) transition on the positive $g$ and two transitions on the negative $g$-side. Our numerical studies suggest a similar phase diagram with one transition on the positive $g$ and \textit{three} transitions on the negative $g$ side. The first transition, for, $g<0$,  is a Lifshitz transition to a critical phase with central charge $c=3/2$. It appears that there is a second transition to another critical phase with central charge $c=2$ upon increasing the negative interaction strength, then a transition to a critical phase with $c=1$, and finally a transition to the gapped phase at very strong negative interactions. Our estimate of these phase transition are shown in Fig.~\ref{fig:schemFD}. An interesting feature of the four-leg ladder is that unlike the chain and the two-leg ladder (and the 2D case), while gapped phases appear at strong coupling, the gaps become smaller upon increasing $|g$ and the $|g|\to \infty$ points are gapless.
\begin{figure}[]
	\includegraphics[width=\columnwidth]{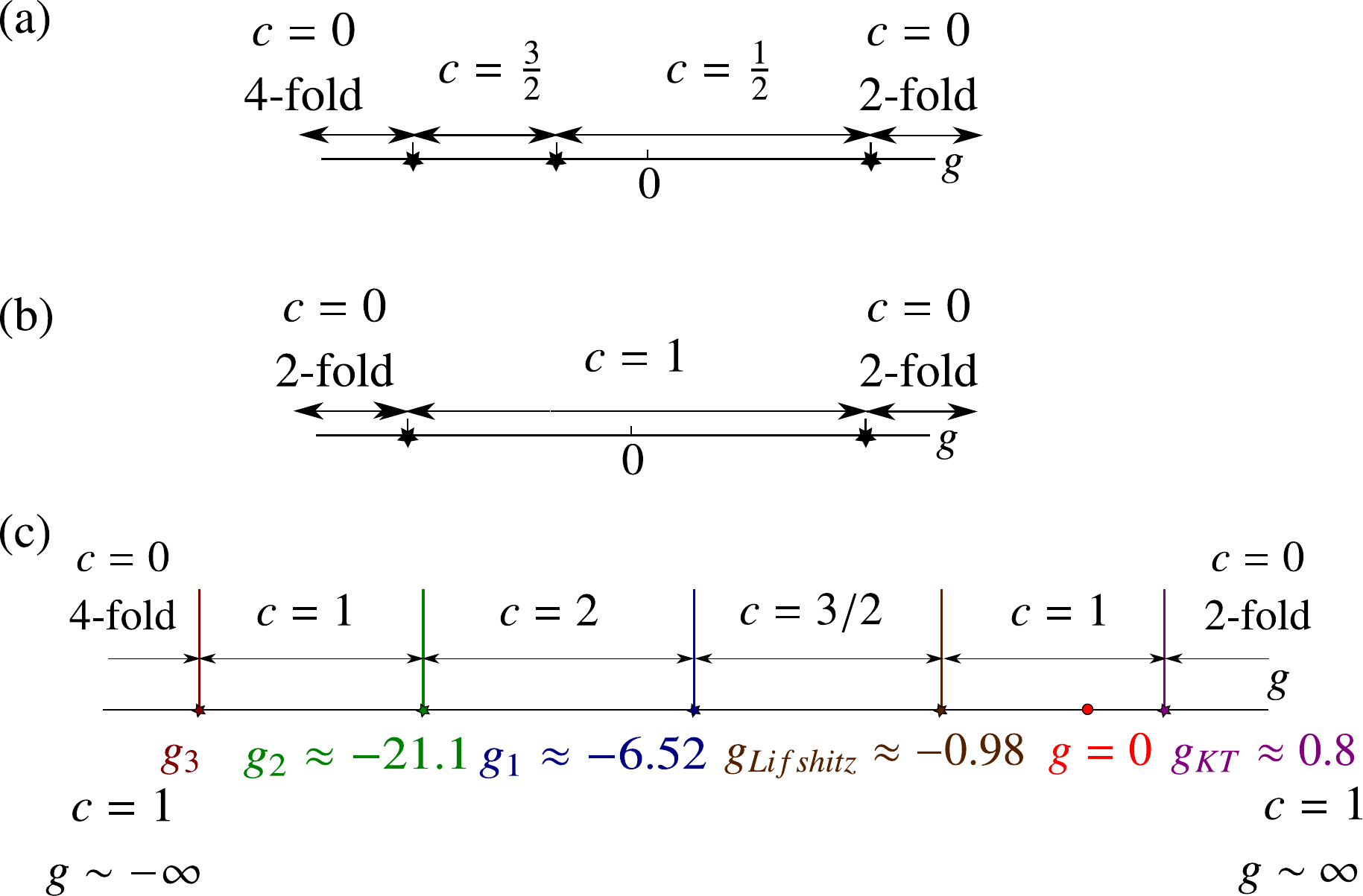}
	\caption{(a) The phase diagram of the Majorana-Hubbard chain. (b) The phase diagram of the two-leg Majorana Hubbard model with peiodic boundary conditions in the $y$ direction. (c) The phase diagram of the four-leg Majorana-Hubbard ladder with periodic boundary conditions in the $y$ direction.  The numerically computed values of the interaction strength $g$ at various phase transitions are shown fr hopping $t=1$. The interaction strength $|g_3|$ is expected to be very large, but we have not been able to determine its value due to the large correlation length of the gapped phase and strong finite-size effects in this regime. \label{fig:schemFD}}
\end{figure}

We combined several numerical diagnostics in determining the phase diagram: a direct calculation of the entanglement entropy, which provides the central charges of the critical phases, calculation of the two-point functions of Majorana operators, which decay as a power-law in critical phases, and direct extrapolation of the spectral gaps. 

\subsection{Entanglement entropy and the central charge}

Our primary tool in determining the phase diagram is based on a calculation of the entanglement entropy between two parts of the ladder using DMRG. We can directly compute the central charge of the system if it is in a gapless phase. Up to subleading terms, which happen to be oscillatory in the present model, we expect the entanglement entropy between a subsystem of length $\ell$ and the remaining subsystem of length $L-\ell$ for a (quasi)-one-dimensional system of length $L$ to be given by
\[
S(\ell)={c\over 6}\ln\left[\sin(\pi \ell/L)\right],
\]
with open boundary conditions in the $x$ direction.
The ground-state entanglement entropy can be measured directly with DRMG and does conform to the predicted form above in several regions of the phase diagram upon canceling out the subleading oscillatory terms using nearest-neighbor averaging, as shown, e.g., in Fig. \ref{fig:ent} for a system of length $L=140$ and $g=-10$.
\begin{figure}[]
	\includegraphics[width=\columnwidth]{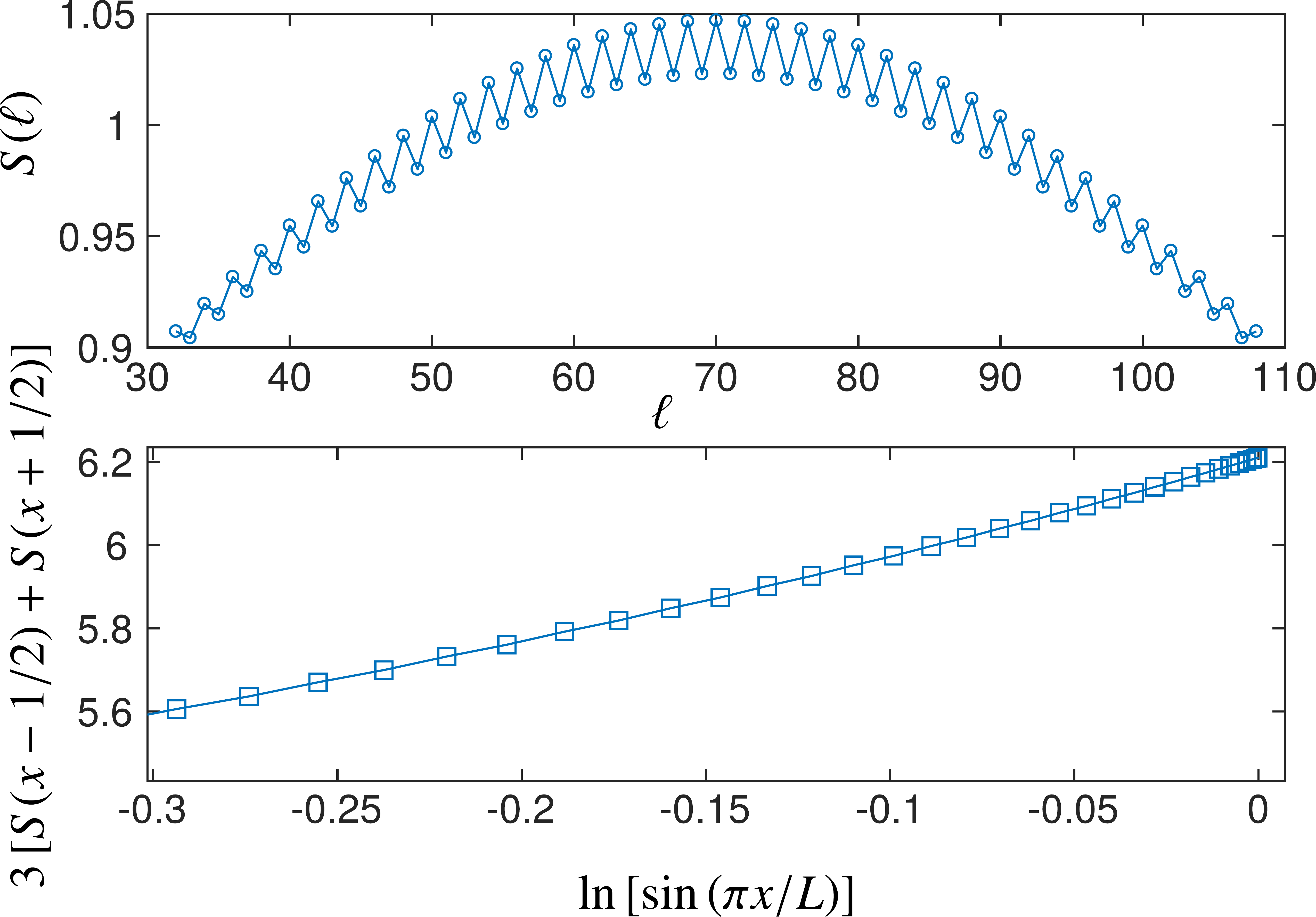}
	\caption{(Top) An example of the entanglement entropy for a system of length $L=140$ and $g=-10$. (Bottom) Averaging the entanglement entropy for two nearest integers $\ell$ and $\ell+1$ (assigning the average to a middle point $x=\ell-1/2$), effectively eliminates the oscillatory subleading terms, allowing us to extract the central charge $c$ from the slop of the resulting linear curve. \label{fig:ent}}
\end{figure}

Using this approach, we can determine the central charge as a function of the interaction strength. We also analyzed the goodness of the fits to the CFT results. A bad fit indicates that we likely have a gapped phase and the $c$ value obtained from fitting $S(\ell)$ should not be trusted. This occurs in the vicinity of several of the phase transitions, e.g., the Lifshitz transition, as well as the gapped phase at large $g$. Our results for the central charge are shown in Fig. \ref{fig:central3}.
%
%

The full phase diagram can be inferred from these plots of the numerically extracted central charge. We know that $g\to\infty$ is a gapped phase and $g=0$ is a gapless Luttinger-liquid phase with $c=1$. Both of these predictions are confirmed by the numerics. A plateau with $c=1$ is clearly observed around $g=0$. Upon increasing $g$, the central charge seems to dip at around $g\approx 0.7$, increase to around $c\approx 1.25$ and then drops to zero. This may indicate a multicritical point at $g\approx 2.0$ with central charge $c=1.25$. However, other diagnostics do not support this picture. It appears that the bumps in the measured central charge is within the gapped phase. We will see this explicitly by calculating the expectation values of the fermionic two-point functions and the extrapolation of the energy gaps. We conclude that for positive $g$ a single transition occurs at around $g\approx 0.8$. On theoretical grounds, we expect this transition to belong to the KT universality class, and the calculations of the Green's function support this prediction. 

On the negative $g$ side, there is strong evidence for a phase transition at around $g\approx -1$. We have strong evidence from the extrapolation of the spectral gap that this transition occurs and is a Lifshitz transition. Stronger interaction strengths give rise to other phases and transitions. Two robust plateaus with central charge $c=2$ and $c=1$ are clearly visible. The values $g_1$ and $g_2$ of $g$, for which these plateaus begin, drift with system size.

To determine and estimated phase diagram, we used a linear extrapolation to estimate the values of $g_1$ and $g_2$ in the thermodynamic limit as shown in Fig. \ref{fig:extrap}.
Our extrapolation of these values strongly suggests that there is a finite phase between the Lifshitz transition and the $c=2$ phase. Although the behavior of the numerically estimated central charge is rather chaotic in this region, it appears that this intermediate phase may have $c=3/2$. In fact at a Lifshitz transition, we expect a species of low-energy fermions to appear and the smallest change can be the appearance of one low-energy Majorana. We leave an in-depth study of the nature of these phases for negative $g$ to future publications. 

Theoretically, we also know that $g\to\infty$ is a gapped phase. In our numerical studies, we were not able to see direct evidence of the gap, which suggests a small gap over a small range of $|t/g|$.

\begin{figure}[]
	\includegraphics[width=\columnwidth]{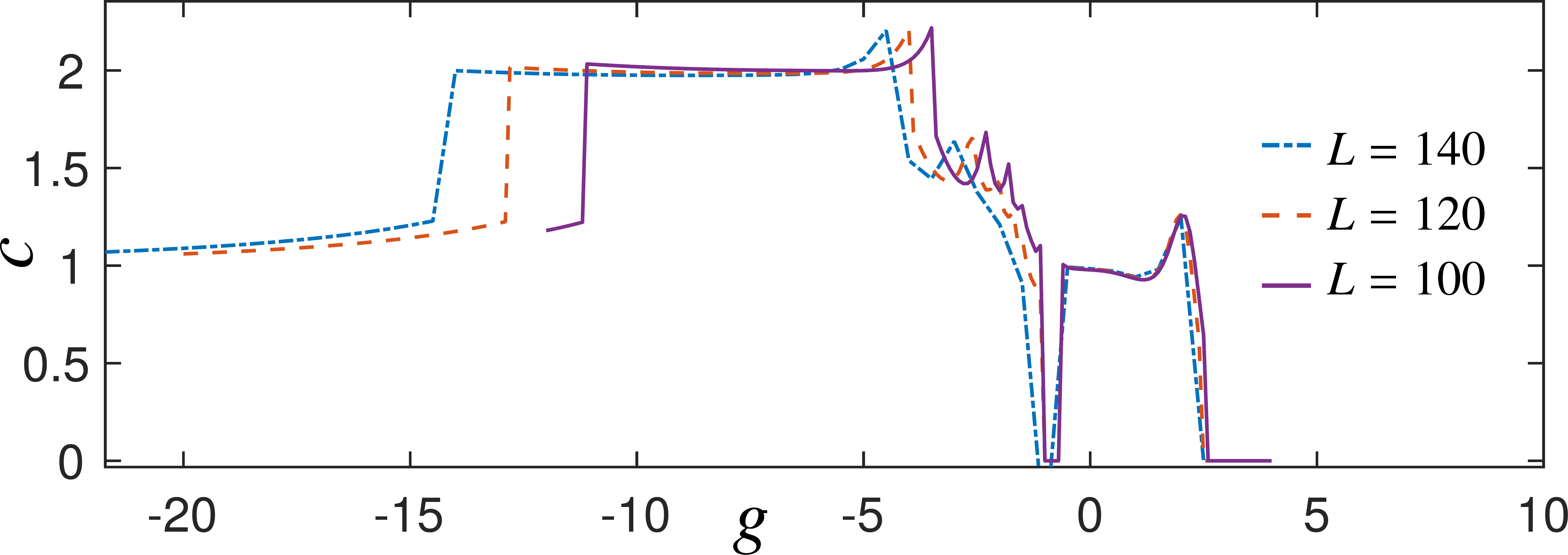}
	\caption{The numerically extracted central charge for several system sizes.\label{fig:central3}}
\end{figure}
\begin{figure}[]
	\includegraphics[width=\columnwidth]{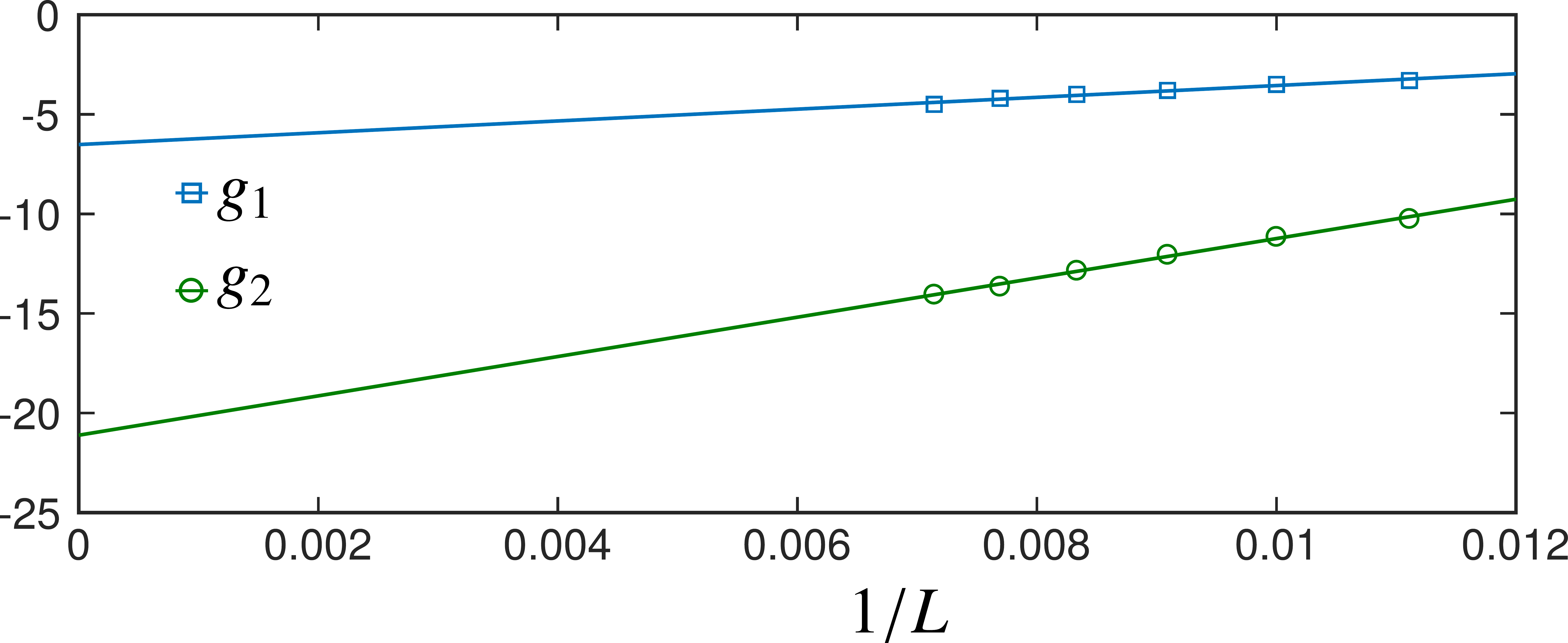}
	\caption{The extrapolation of $g_1$ and $g_2$, using $L=90, 100, \dots, 140$ and a linear fit (to $1/L$) of of the values of $g_{1,2}$, for which the central charge jumps in finite-size numerical results, as seen in Fig.~\ref{fig:central3}.\label{fig:extrap}}
\end{figure}

\subsection{Majorana Green's function and the KT transition}
DMRG allows us to also compute the ground-state expectation values of various operators. In this section, we focus on 
\begin{equation}
G(x)=i\langle \gamma_{m,n}\gamma_{m,n+x}\rangle.
\end{equation}
As we have periodic boundary condition in the $y$ direction, the Green's function is independent of $m$. With periodic and antiperiodic boundary conditions in the $x$ direction, it is also independent of $n$. We have found that antiperiodic boundary conditions suppress the subleading terms and allow us to extract the universal behavior of $G(x)$ from the numerics more conveniently. If the system is in a Luttinger-liquid phase with Luttinger parameter $K$, we expect 
\begin{equation}
G(x)\sim x^{(K+1/K)/2}.
\end{equation}
In a finite system, we can replace $x$ with ${L\over \pi}\sin{\pi x/L}$. In Fig. \ref{green}, we show the behavior of $G(x)$ on the positive $g$ side. For small $g$, we observe the expected power-law decay, with an increasing exponent as we increase $g$. The behavior transitions to a faster decay for larger $g$ around $g=0.8$ with $K\approx 1/2$. This is consistent with a KT transition, as analyzed in Sec. VIII. The bump in the central charge around $g=2$ seems to be an artifact as there is evidence from the Green's function that we have already entered a gapped phase at this value of $g$.

\begin{figure}[]
	\includegraphics[width=\columnwidth]{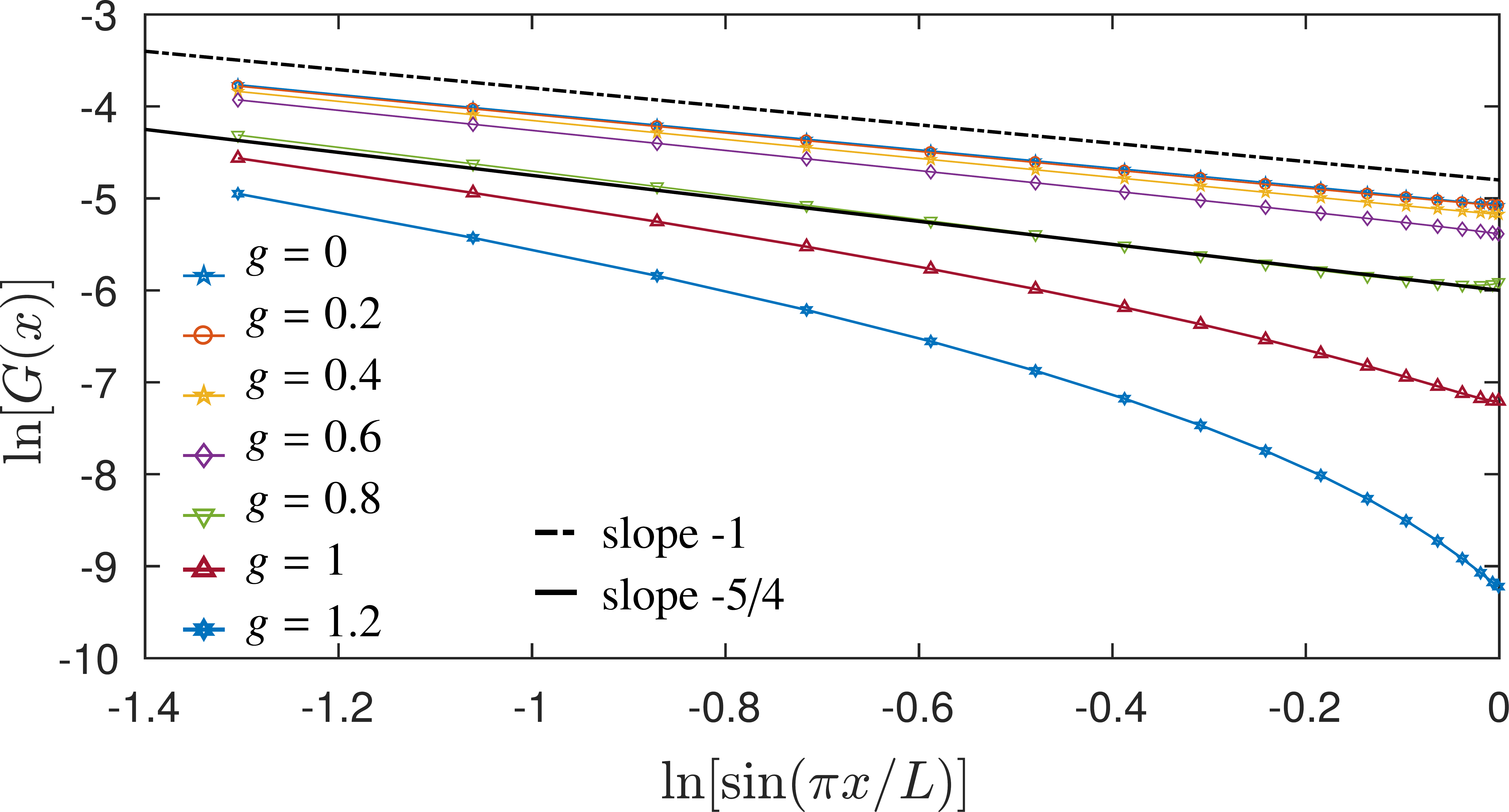}
	\caption{The numerically computed Green's function for positive $g$ for a system of length $L=80$. For $g=0$ and small values of $g$, the expected power law is observed. When the exponent reaches ${1\over 2}(K+1/K)\approx 5/4$ at around $g\approx 0.8$, which corresponds $K=1/2$, the behavior of the correlation function shift to a decay faster than power law. .\label{green}}
\end{figure}

A curious feature of the central charge data is that for $0.8<g<2$, the entanglement entropy looks similar to that of a critical phase and a robust peak is observed at around $g=2$. The behavior of the correlation functions are, however, suggestive of a gapped phase. It is possible that a large correlation length in finite systems gives the illusion of criticality in the behavior of the entanglement entropy. To investigate this issue further, we directly calculated the energy gap $\Delta E$ (with antiperiodic boundary conditions and in the even fermion parity sector).

As shown in Fig.~\ref{fig:extrapE}, the results indicate that the gap at $g=0.8$ likely extrapolates to a finite value in the thermodynamic limit (we used a fit to a second-order polynomial of $1/L$). For a slightly larger $g=1$, there is strong evidence of a finite energy gap. This suggests that the critical behavior of the entanglement entropy is most likely an artifact.

\begin{figure}[]
	\includegraphics[width=\columnwidth]{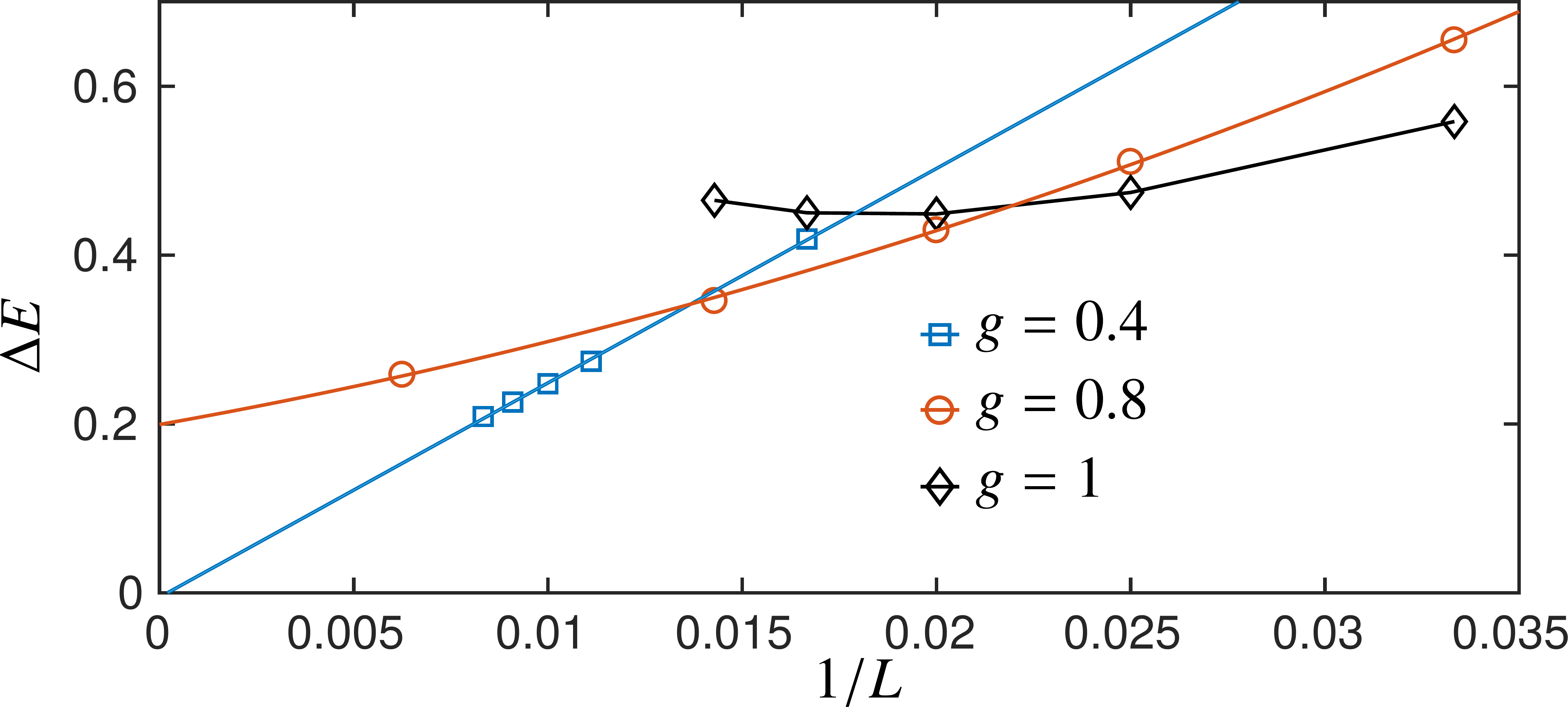}
	\caption{The energy gap $\Delta E$ extrapolates to zero (using a linear fit to $1/L$ for $g=0.4$. Upon increasing the interaction strength, the linear fits begin to fail but a quadratic fit to $1/L$ is suggestive of extrapolation to a finite value at $g=0.8$, near which we expect a transition from the behavior of the Green's function. For a slightly larger $g=1$, there is clear evidence of an energy gap in the thermodynamic limit.\label{fig:extrapE}}
\end{figure}

\subsection{Lifshitz transition and the Luttinger-liquid velocity}
We now focus on the negative $g$ side. The first transition out of the Luttinger liquid phase is easier to understand. We claim this transition is a Lifshitz transition analogous to the Majorana chain.
Direct evidence is provided by extracting the velocity in the Luttinger-liquid phase and observing that it extrapolated to zero, signaling the emergence of a dynamical exponent $z>1$ at the transition. In the Luttinger-liquid phase the energy gaps scale as $1/L$. In particular we define the velocity as the coefficient $v$ in 
\begin{equation}
\Delta E=E^{\rm even}_1-E^{\rm even}_0=2\pi x v/L,
\end{equation}
where $\Delta E$ is defined as the gap from the ground state to the first excited state in the even fermion party sector, and $x$ is a universal constant of order unity (the scaling dimension of the operator corresponding to this energy level). As shown, in Fig. \ref{fig:gap}, this behavior is confirmed in the numerics, allowing us to extract the coefficient $\tilde v=2\pi x v$ from a linear fit.

\begin{figure}[]
	\includegraphics[width=0.8\columnwidth]{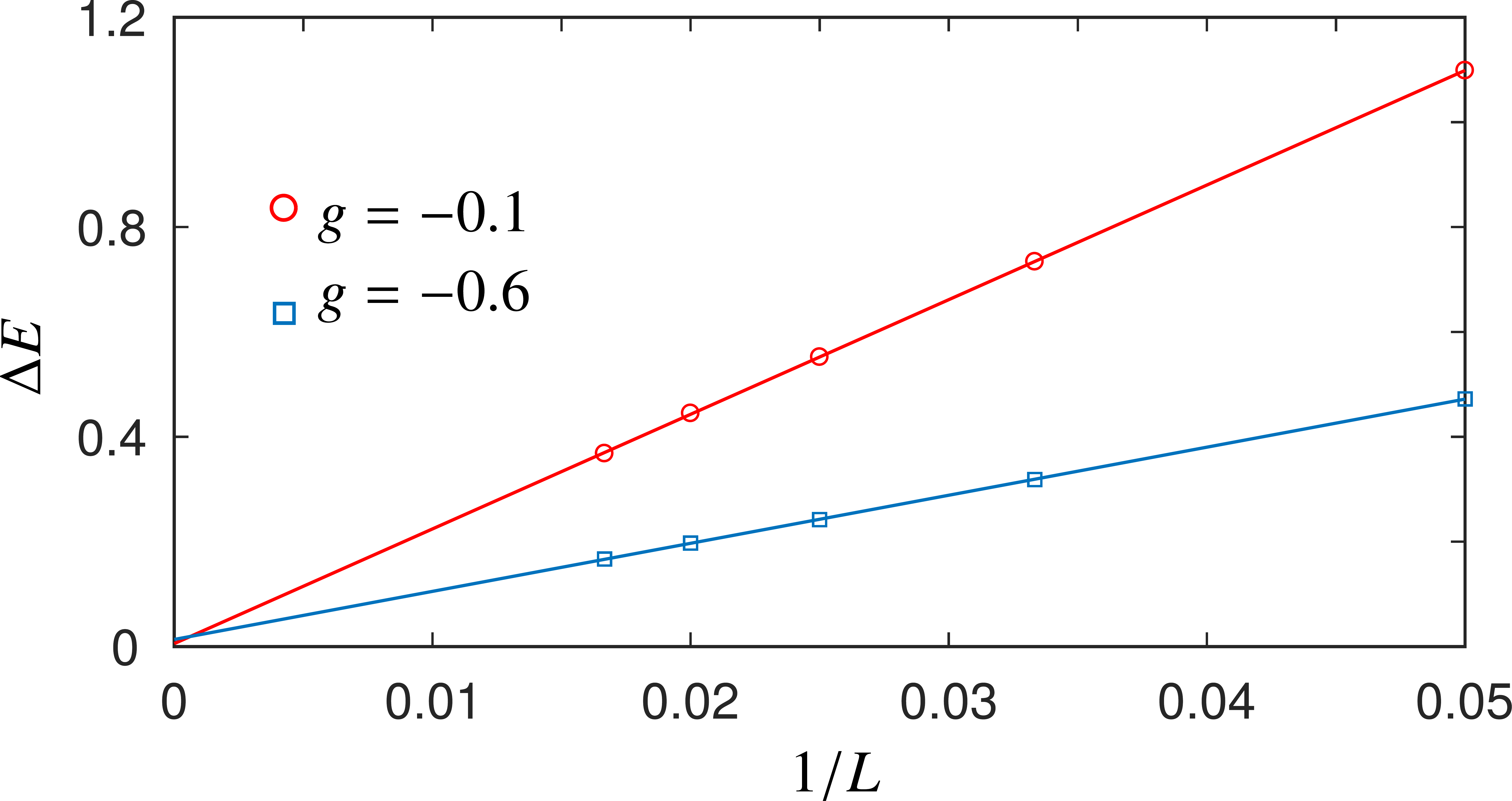}
	\caption{The energy gaps in the even fermion parity sector as a function of $1/L$ for $g$ in the LL phase. }\label{fig:gap}
\end{figure}
Extracting $\tilde v\propto v$ and plotting it as a function of $g$ allows us to identify the location of the Lifshitz transition, see Fig. \ref{fig:velocity}. We note that the simple linear dependence disappears as we move past the transition, and an intricate dependence on system size appears similar to the $c=3/2$ phase of the chain. 
\begin{figure}[]
	\includegraphics[width=\columnwidth]{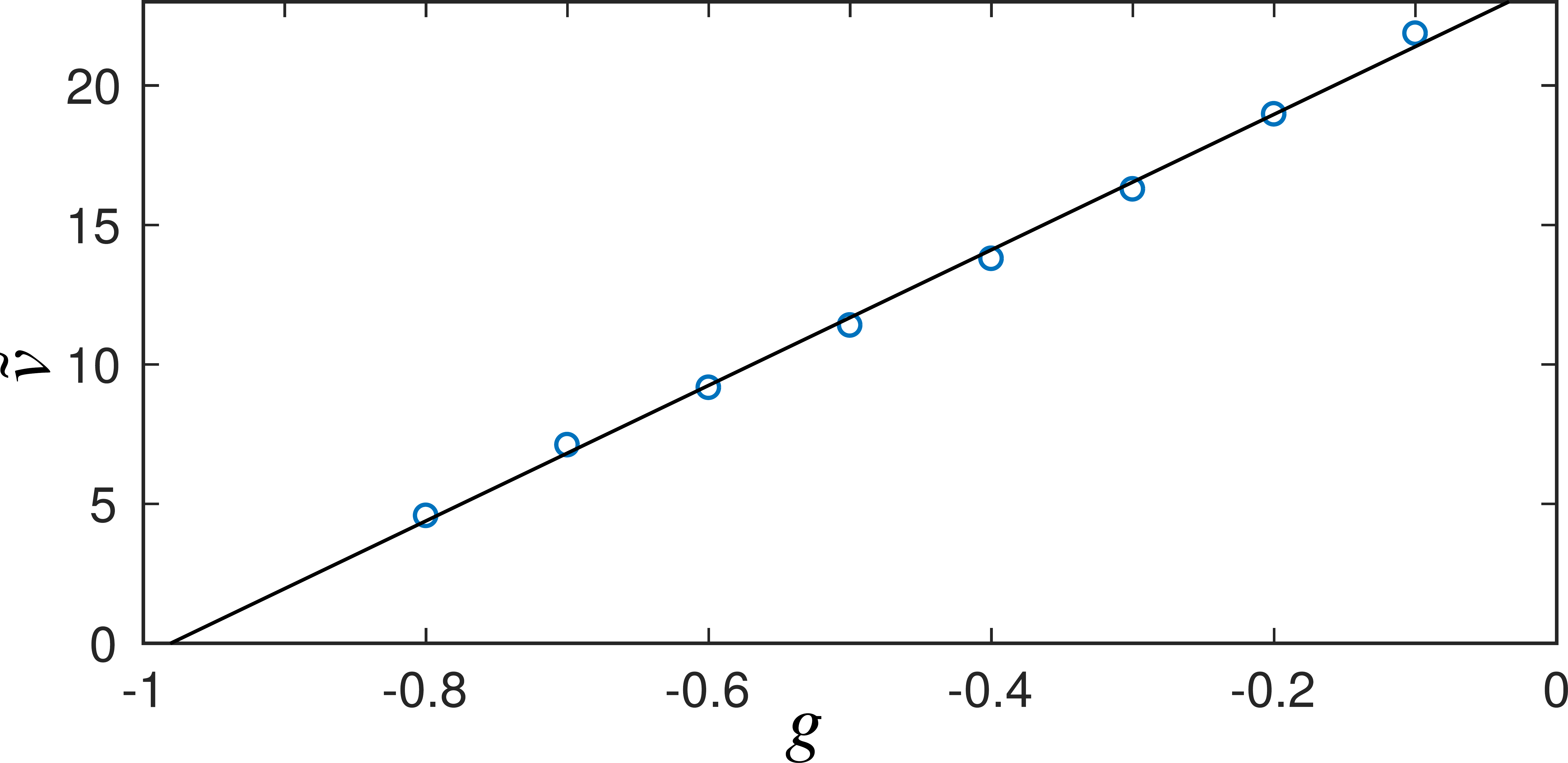}
	\caption{The behavior of the velocity as a function of $g$ signaling a Lifshitz transition at around $g=-0.98$, where the velocity vanishes, in agreement with the central charge results. }\label{fig:velocity}
	\end{figure}

  \section{Conclusions}
 We have studied the two-leg and four-leg Majorana-Hubbard model.  The behavior is largely consistent with our previous mean-field predictions for the 
 2D case.  For PCB, a massless phase occurs at sufficiently weak coupling of either sign. As $|g|$ increases, transitions occur to broken symmetry phases with 
 neighboring Majorana fermions combining to form Dirac fermion levels which tend to be empty or filled. While in the 2D case these 
 can occur on vertical or horizontal bonds, for ladders, they only occur on vertical bonds, reducing the number of ground states. We also 
 found, in the four-leg case at large negative $g/t$, that the dimer order has a larger unit cell than predicted by our naive mean-field theory, with 
 the Dirac levels alternating filled and empty in both horizontal and vertical directions. 
 
\begin{acknowledgments}
The authors would like to thank Marcel Franz, Tarun Tummuru, and Timothy Hsieh for helpful discussions. 
This research is supported by NSERC Discovery Grant 04033-2016 and by the Canadian Institute for Advanced Research. 
\end{acknowledgments}
\appendix

 \section{Spin Ladder Mapping}
 We first rewrite the Hamiltonian in terms of momentum $0$ and $\pi$ channels, with periodic boundary conditions. For $g=0$, the momentum $0$ channel is gapless 
 and the momentum $\pi$ channel is gapped. While the two channels are decoupled in the noninteracting model, the interaction term couples them together 
 (while also introducing interactions within each channel).  It is then convenient to attempt to integrate out the gapped $\pi$ channel to get an effective 
 Hamiltonian for the gapless $0$ channel in order to study the phase transitions. 
 
 The momentum $0$ and $\pi$ channels are defined by:  
 \bea && \gamma^e_{m,0}\equiv {1\over \sqrt{2}}(\gamma_{m,0}+\gamma_{m,2}),\ \  \gamma^o_{m,0}\equiv {1\over \sqrt{2}}(\gamma_{m,1}+\gamma_{m,3})\nonumber \\
 &&\gamma^e_{m,\pi}\equiv {1\over \sqrt{2}}(\gamma_{m,0}-\gamma_{m,2}),\ \ \gamma^o_{m,\pi}\equiv {1\over \sqrt{2}}(\gamma_{m,1}-\gamma_{m,3}).\nonumber
 \eea
 Solving:
 \bea &&\gamma_{m,0}={1\over \sqrt{2}}(\gamma^e_{m,0}+\gamma^e_{m,\pi}),\ \  \gamma_{m,1}={1\over \sqrt{2}}(\gamma^o_{m,0}+\gamma^o_{m,\pi})\nonumber \\
 &&\gamma_{m,2}={1\over \sqrt{2}}(\gamma^e_{m,0}-\gamma^e_{m,\pi}),\ \ \gamma_{m,3}={1\over \sqrt{2}}(\gamma^o_{m,0}-\gamma^o_{m,\pi}).\nonumber
 \eea

\begin{widetext} 
The vertical hopping term contains
 \bea &&\sum_{n=0}^3\gamma_{m,n}\gamma_{m,n+1}={1\over 2}[(\gamma^e_{m,0}+\gamma^e_{m,\pi})(\gamma^o_{m,0}+\gamma^o_{m,\pi})+(\gamma^o_{m,0}+\gamma^o_{m,\pi})(\gamma^e_{m,0}-\gamma^e_{m,\pi})
 \nonumber \\
 &&+(\gamma^e_{m,0}-\gamma^e_{m,\pi})(\gamma^o_{m,0}-\gamma^o_{m,\pi})+(\gamma^o_{m,0}-\gamma^o_{m,\pi})(\gamma^e_{m,0}+\gamma^e_{m,\pi})]=2\gamma^e_{m,\pi}\gamma^o_{m,\pi}.
\label{VHT} \eea
 This term produces a gap in the $\pi$ channel only. The horizontal hopping term contains:
 \bea && \gamma_{m,0}\gamma_{m+1,0}-\gamma_{m,1}\gamma_{m+1,1}+\gamma_{m,2}\gamma_{m+1,2}-\gamma_{m,3}\gamma_{m+1,3}\nonumber \\
 &&={1\over 2}[(\gamma^e_{m,0}+\gamma^e_{m,\pi})(\gamma^e_{m+1,0}+\gamma^e_{m+1,\pi})-(\gamma^o_{m,0}+\gamma^o_{m,\pi})(\gamma^o_{m+1,0}+\gamma^o_{m+1,\pi})\nonumber \\
 &&+(\gamma^e_{m,0}-\gamma^e_{m,\pi})(\gamma^e_{m+1,0}-\gamma^e_{m+1,\pi})-(\gamma^o_{m,0}-\gamma^o_{m,\pi})(\gamma^o_{m+1,0}-\gamma^o_{m+1,\pi})]\nonumber \\
 &&=\gamma^e_{m,0}\gamma^e_{m+1,0}+\gamma^e_{m,\pi}\gamma^e_{m+1,\pi}-\gamma^o_{m,0}\gamma^o_{m+1,0}-\gamma^o_{m,\pi}\gamma^o_{m+1,\pi}
 \eea
 Thus the total hopping term is:
 \be H_0=it\sum_m[\gamma^e_{m,0}\gamma^e_{m+1,0}+\gamma^e_{m,\pi}\gamma^e_{m+1,\pi}-\gamma^o_{m,0}\gamma^o_{m+1,0}-\gamma^o_{m,\pi}\gamma^o_{m+1,\pi}+2\gamma^e_{m,\pi}\gamma^o_{m,\pi}].
 \ee
 We define Dirac modes
 \be c_{m,0}\equiv {\gamma^e_{m,0}+i\gamma^0_{m,0}\over 2},\ \  c_{m,\pi}\equiv {\gamma^e_{m,\pi}+i\gamma^0_{m,\pi}\over 2}
 \ee
 and use
 \be \gamma^e_m\gamma^e_{m+1}-\gamma^o_m\gamma^o_{m+1}=(c_m^\dagger +c_m)(c_{m+1}^\dagger +c_{m+1})+(c_m^\dagger -c_m)(c_{m+1}^\dagger -c_{m+1})
 =2(c_mc_{m+1}-c^\dagger_{m+1}c^\dagger_m),\ee
 \be H_0=2t\sum_m[i(c_{m,0}c_{m+1,0}-c^\dagger_{m+1,0}c^\dagger_{m,0})+i(c_{m,\pi}c_{m+1,\pi}-c^\dagger_{m+1,\pi}c^\dagger_{m,\pi})+2(2c^\dagger_{m,\pi}c_{m,\pi}-1)].\ee
 To get this into a more standard form, we make the transformation:
 
 \bea c_{2m,k}&\to& (-1)^mc_{2m,k},\nonumber \\
 c_{2m+1,k}&\to& -i(-1)^mc_{2m+1,k}^\dagger \label{trans}\eea
 for both $k=0$ and $\pi$ modes, giving
 \be H_0=2t\sum_m[-(c^\dagger_{m,0}c_{m+1,0}+c^\dagger_{m,\pi}c_{m+1,\pi}+{\rm H.c.})+(-1)^m2(2c^\dagger_{m,\pi}c_{m,\pi}-1)].\label{eq:tran}\ee
 We see that the $k=0$ mode is in a gapless XY phase, while the $k=\pi$ mode is gapped due to a staggered field in the spin representation. 
 
 We now perform similar steps to write the interaction in terms of the Dirac fermions.
 \bea H_{int}&=&g\sum_m[\gamma_{m,0}\gamma_{m+1,0}\gamma_{m+1,1}\gamma_{m,1}+\gamma_{m,1}\gamma_{m+1,1}\gamma_{m+1,2}\gamma_{m,2}+
 \gamma_{m,2}\gamma_{m+1,2}\gamma_{m+1,3}\gamma_{m,3}+\gamma_{m,3}\gamma_{m+1,3}\gamma_{m+1,0}\gamma_{m,0}]\nonumber \\
 &=&{g\over 4}\sum_m[(\gamma^e_{m,0}+\gamma^e_{m,\pi})(\gamma^e_{m+1,0}+\gamma^e_{m+1,\pi})(\gamma^o_{m+1,0}+\gamma^o_{m+1,\pi})(\gamma^o_{m,0}+\gamma^o_{m,\pi})\nonumber \\
 &+&(\gamma^o_{m,0}+\gamma^o_{m,\pi})(\gamma^o_{m+1,0}+\gamma^o_{m+1,\pi})(\gamma^e_{m+1,0}-\gamma^e_{m+1,\pi})(\gamma^e_{m,0}-\gamma^e_{m,\pi})\nonumber \\
 &+&(\gamma^e_{m,0}-\gamma^e_{m,\pi})(\gamma^e_{m+1,0}-\gamma^e_{m+1,\pi})(\gamma^o_{m+1,0}-\gamma^o_{m+1,\pi})(\gamma^o_{m,0}-\gamma^o_{m,\pi})\nonumber \\
 &+&(\gamma^o_{m,0}-\gamma^o_{m,\pi})(\gamma^o_{m+1,0}-\gamma^o_{m+1,\pi})(\gamma^e_{m+1,0}+\gamma^e_{m+1,\pi})(\gamma^e_{m,0}+\gamma^e_{m,\pi})].\nonumber \\
 &=&g\sum_m[\gamma^e_{m,0}\gamma^e_{m+1,0}\gamma^o_{m+1,0}\gamma^o_{m,0}+\gamma^e_{m,\pi}\gamma^e_{m+1,\pi}\gamma^o_{m+1,\pi}\gamma^o_{m,\pi}\nonumber \\
 &+&\gamma^e_{m,0}\gamma^e_{m+1,0}\gamma^o_{m+1,\pi}\gamma^o_{m,\pi}+\gamma^o_{m,0}\gamma^o_{m+1,0}\gamma^e_{m+1,\pi}\gamma^e_{m,\pi}]\nonumber \\
 &=&g\sum_m[-(2c^\dagger_{m,0}c_{m,0}-1)(2c^\dagger_{m+1,0}c_{m+1,0}-1)-(2c^\dagger_{m,\pi}c_{m,\pi}-1)(2c^\dagger_{m+1,\pi}c_{m+1,\pi}-1)\nonumber \\
 &-& (c^\dagger_{m,0}+c_{m,0})(c^\dagger_{m+1,0}+c_{m+1,0})(c^\dagger_{m+1,\pi}-c_{m+1,\pi})(c^\dagger_{m,\pi}-c_{m,\pi})\nonumber \\
 &-& (c^\dagger_{m,\pi}+c_{m,\pi})(c^\dagger_{m+1,\pi}+c_{m+1,\pi})(c^\dagger_{m+1,0}-c_{m+1,0})(c^\dagger_{m,0}-c_{m,0})].
 \eea
 Next, we transform the $c$ operators as in Eq. (\ref{trans}), giving
 \bea H_{int}&=&g\sum_m[(2c^\dagger_{m,0}c_{m,0}-1)(2c^\dagger_{m+1,0}c_{m+1,0}-1)+(2c^\dagger_{m,\pi}c_{m,\pi}-1)(2c^\dagger_{m+1,\pi}c_{m+1,\pi}-1)\nonumber \\
 &-&  (c^\dagger_{m,0}+c_{m,0})(c^\dagger_{m+1,0}-c_{m+1,0})(c^\dagger_{m+1,\pi}+c_{m+1,\pi})(c^\dagger_{m,\pi}-c_{m,\pi})\nonumber \\
 &-& (c^\dagger_{m,\pi}+c_{m,\pi})(c^\dagger_{m+1,\pi}-c_{m+1,\pi})(c^\dagger_{m+1,0}+c_{m+1,0})(c^\dagger_{m,0}-c_{m,0})].\label{Hintc}
 \eea
 Ignoring interleg coupling this is just two copies of the spinless fermion model with nearest-neighbor interactions and a staggered potential for the $\pi$ sector.  The $0$ sector remains gapless for $|g|<t$.

 Next we make a Jordan-Wigner transformation to write the Hamiltonian in terms of spin-$1\over 2$ degrees of freedom.
 \bea c_{m,0}&=&(\prod_{m'<m}\sigma^z_{m',0}\sigma^z_{m',\pi})\sigma^-_{m,0}\nonumber \\
 c_{m\pi}&=&(\prod_{m'<m}\sigma^z_{m',0}\sigma^z_{m',\pi})\sigma^z_{m,0}\sigma^-_{m\pi}\nonumber \\
 2c^\dagger_{m,k}c_{m,k}-1&=&\sigma^z_{m,k}\eea
 for $k=0$ and $\pi$. The hopping terms become:
 \be H_0=t\sum_m[(\sigma^+_{m,0}\sigma^z_{m,\pi}\sigma^-_{m+1,0}+\sigma^-_{m,0}\sigma^z_{m,\pi}\sigma^+_{m+1,0})+(\sigma^+_{m,\pi}\sigma^z_{m+1,0}\sigma^-_{m+1,\pi}+\sigma^-_{m,\pi}\sigma^z_{m+1,0}\sigma^+_{m+1,\pi})+2(-1)^m\sigma^z_{m,\pi}].\label{H0s}
 \ee
To transform the interaction term note that
\bea  && (c^\dagger_{m,0}+c_{m,0})(c^\dagger_{m,\pi}-c_{m,\pi})(c^\dagger_{m+1,0}+c_{m+1,0})(c^\dagger_{m+1,\pi}-c_{m+1,\pi})\nonumber \\
&&\to (\sigma^+_{m,0}+\sigma^-_{m,0})\sigma^z_{m,0}(\sigma^+_{m,\pi}-\sigma^-_{m,\pi})(\sigma^+_{m+1,0}+\sigma^-_{m+1,0})\sigma^z_{m+1,0}(\sigma^+_{m+1,\pi}-\sigma^-_{m+1,\pi})\nonumber \\
&&= (-\sigma^+_{m,0}+\sigma^-_{m,0})(\sigma^+_{m,\pi}-\sigma^-_{m,\pi})(-\sigma^+_{m+1,0}+\sigma^-_{m+1,0})(\sigma^+_{m+1,\pi}-\sigma^-_{m+1,\pi})\nonumber \\
&&=\sigma^y_{m,0}\sigma^y_{m,\pi}\sigma^y_{m+1,0}\sigma^y_{m+1,\pi}.
\eea
 \be H_{int}=g\sum_m[\sigma^z_{m,0}\sigma^z_{m+1,0}+\sigma^z_{m,\pi}\sigma^z_{m+1,\pi}-\sigma^y_{m,0}\sigma^y_{m+1,0}\sigma^y_{m,\pi}\sigma^y_{m+1,\pi}
 -\sigma^x_{m,0}\sigma^x_{m+1,0}\sigma^x_{m,\pi}\sigma^x_{m+1,\pi}].\label{Hint}
 \ee
 We effectively get a 2-leg spin ladder with the legs corresponding to $k=0$ and $k=\pi$ modes.  We have an unusual interleg 4-spin coupling which breaks the U(1) symmetry of the decoupled legs. 
 Ignoring $H_0$, we can find the exact ground state of $H_{int}$ as we saw earlier in the Majorana basis. Here we confirm that we can also solve it exactly in the spin basis. 
 The rung fermion parity is
\be i\gamma_0\gamma_1+i\gamma_2\gamma_3=i\gamma^e_0\gamma^o_0+i\gamma^e_{\pi}\gamma^o_{\pi}\to \sigma^z_0+\sigma^z_{\pi}.\ee
Equation (\ref{Hint}) preserves rung parity since $\sigma^{x/y}_{m,0}\sigma^{x/y}_{m,\pi}$ changes the value of 
\be \sigma^z_m\equiv \sigma^z_{m,0}+\sigma^z_{m,\pi}\ee
by either $0$ or $\pm 4$. $\sigma^z_m$ takes the values $\pm 2$ in one sector and $0$ in the other. In the  $\pm 2$ sector, we may replace:
\be \sigma^x_0\sigma^x_\pi \to (\sigma^+_0\sigma^+_\pi + {\rm H.c.}),\ \ 
\sigma^y_0\sigma^y_\pi \to -(\sigma^+_0\sigma^+_\pi + {\rm H.c.}).\ee
While $\sigma^x$ simply flips the spin, we have $\sigma^y|\uparrow\rangle=i|\downarrow\rangle$ and $\sigma^y|\downarrow\rangle=-i|\uparrow\rangle$. The $\pm 2$ sector is spanned by $|\uparrow\uparrow\rangle$ and $|\downarrow\downarrow\rangle$, and the minus sign in the above expression originates from $(\pm i)^2$ associated with the action of $\sigma^y_0\sigma^y_\pi $.
In the $0$ sector, we may replace:
\be \sigma^x_0\sigma^x_\pi \to (\sigma^+_0\sigma^-_\pi + {\rm H.c.}), \ \ 
\sigma^y_0\sigma^y_\pi \to (\sigma^+_0\sigma^-_\pi + {\rm H.c.}).\ee
This sector is spanned by $|\uparrow\downarrow\rangle$ and $|\downarrow\uparrow\rangle$ so the $\sigma^y_0\sigma^y_\pi $ term gives a factor of $-(i)^2=1$. If two neighboring sites are in opposite sectors the corresponding term in the Hamiltonian vanishes. For the 
\be \sigma^z_{m,0}\sigma^z_{m+1,0}+\sigma^z_{m,\pi}\sigma^z_{m+1,\pi}\ee
terms this follows because, for example, if $m$ is in the $0$ sector and $m+1$ is in the $\pm 2$ sector then 
$\sigma^z_{m,0}=-\sigma^z_{m,\pi}$ and $\sigma^z_{m+1,0}=\sigma^z_{m+1,\pi}$. For the XXYY terms this follows because
\bea &&\sigma^x_{m,0}\sigma^x_{m,\pi}\sigma^x_{m+1,0}\sigma^x_{m+1,\pi}+
 \sigma^y_{m,0}\sigma^y_{m,\pi}\sigma^y_{m+1,0}\sigma^y_{m+1,\pi}\nonumber \\
&&\to 
(\sigma^+_{m,0}\sigma^-_{m,\pi}+h.c.)(\sigma^+_{m+1,0}\sigma^+_{m+1,\pi}+h.c.)
-(\sigma^+_{m,0}\sigma^-_{m,\pi}+h.c.)(\sigma^+_{m+1,0}\sigma^+_{m+1,\pi}+h.c.)=0.
\eea

\end{widetext}
Thus we may assume that each rung is in the same fermion parity sector. There are then only 2 states on each rung. In the 
$0$ sector we may label them:
\be |\uparrow \rangle \equiv |\uparrow ,\downarrow \rangle ,\ \  |\downarrow \rangle \equiv |\downarrow ,\uparrow \rangle ,\ee
and in the $\pm 2$ sector:
\be |\uparrow \rangle \equiv |\uparrow ,\uparrow \rangle ,\ \  |\downarrow \rangle \equiv |\downarrow ,\downarrow \rangle .\ee
Here the first arrow refers to the $0$ leg and second arrow to the $\pi$ leg. 
In the $\pm 2$ sector, for example, we may replace
\bea \sigma^x_{m,0}\sigma^x_{m,\pi}&\to& \sigma^+_{m,0}\sigma^+_{m,\pi}+h.c. \to \sigma^+_m+\sigma^-_m=\sigma^x_m\nonumber \\
 \sigma^y_{m,0}\sigma^y_{m,\pi}&\to&- \sigma^+_{m,0}\sigma^+_{m,\pi}+h.c. \to -\sigma^+_m-\sigma^-_m=-\sigma^x_m\nonumber
\eea
 Similarly, we may replace
 \be \sigma^z_{m,0/\pi}\to \sigma^z_m\ee
 Thus:
 \be H_{int}\to -2g\sum_m[\sigma^z_m\sigma^z_{m+1}+\sigma^x_m\sigma^x_{m+1}]\ee
 Now consider adding a small hopping term.  The horizontal hopping term changes the fermion parity on a pair of channels so it can be ignored in lowest order perturbation theory. The vertical hopping term becomes:
 \be H_0\approx \pm t\sum_m(-1)^m\sigma^z_m\ee 
 where the $+$ factor occurs in the $\pm 2$ sector and the $-$ factor in the $0$ sector.  Thus we have reproduced the results of Sec. IV in the spin basis. 
 
\bibliography{maj}

\end{document}